\documentclass{aa}  
 \usepackage{natbib}  
  \bibpunct{(}{)}{;}{a}{}{,}             
 \usepackage{graphicx}
\usepackage{multirow}
\usepackage{txfonts}

 \usepackage[breaklinks=true]{hyperref} 
 \hypersetup{colorlinks=true,urlcolor=blue,citecolor=blue,pdfborder= 0 0 0}

\def \PKS {PKS~2155-304}

\begin{document} 

\title{To be or not to be: hot WHIM absorption in the blazar \PKS\ sight line?}

\author{J.   Nevalainen\inst{1}\fnmsep\thanks{jukka@to.ee}
          \and
          E. Tempel\inst{1,2}
          \and
            J. Ahoranta\inst{3}
          \and
          L. J. Liivam{\"{a}}gi\inst{1}
           \and
           M. Bonamente\inst{4}
           \and
           E. Tilton\inst{5}
           \and
           J. Kaastra\inst{6,7}
           \and
           T. Fang\inst{8}
          \and
          P. Hein{\"{a}}m{\"{a}}ki\inst{9}
          \and
          E. Saar\inst{1}
          \and
          A. Finoguenov\inst{3}
 }
\institute{Tartu Observatory, University of Tartu, 61602 T{\~{o}}ravere, Tartumaa, Estonia 
  \and
Leibniz-Institut f\"ur Astrophysik Potsdam (AIP), An der Sternwarte 16, 14482 Potsdam, Germany
\and
Department of Physics, University of Helsinki, Gustaf H{\"{a}}llstr{\"{o}}min katu 2a, 00014, Helsinki, Finland
\and
University of Alabama in Huntsville, Huntsville, AL 35899, USA
\and
Department of Physics \&  Astronomy, Regis University, Denver, CO 80221, USA
\and
SRON Netherlands Institute for Space Research, Sorbonnelaan 2, 3584 CA Utrecht, the Netherlands
\and
Leiden Observatory, Leiden University, Niels Bohrweg 2, 2300 RA Leiden, the Netherlands
\and
Xiamen University, No. 422, Siming South Road, Xiamen, Fujian, China
\and
Tuorla Observatory, V{\"{a}}is{\"{a}}l{\"{a}}ntie 20, FI-21500  Piikki{\"{o}}, Finland
}

   \date{Received ; accepted}

   \abstract{The cosmological missing baryons at z$<$1 most likely hide in the hot (T~$\gtrsim10^{5.5}$~K) phase of the  Warm Hot Intergalactic Medium (WHIM).
     While the hot WHIM is hard to detect due to its high ionisation level, the warm (T~$\lesssim10^{5.5}$~K) phase of the WHIM has been very robustly detected in the FUV band. We adopted the assumption that the hot and warm WHIM phases are co-located and thus used the FUV-detected warm WHIM as a tracer for the cosmologically interesting hot WHIM. We utilised the assumption by performing an X-ray follow-up in the sight line of a blazar  \PKS\ at the redshifts where previous FUV measurements of \ion{O}{VI}, \ion{Si}{IV} and BLA absorption have indicated the existence of the warm WHIM. We looked for the \ion{O}{VII}~He$\alpha$ and \ion{O}{VIII}~Ly$\alpha$ absorption lines, the most likely hot WHIM tracers. Despite of the very large exposure time ($\approx$~1~Ms), the XMM-Newton/RGS1 data yielded no significant detection 
which corresponds to 
upper limits of $\log{N(\ion{O}{VII}({\rm cm}^{-2}))}~\le~14.5-15.2$ and $\log{N(\ion{O}{VIII}({\rm cm}^{-2}))}~\le~14.9-15.2$.
An analysis of LETG/HRC data yielded consistent results. However, the LETG/ACIS data yielded a detection of an absorption line - like feature at $\lambda~\approx$~20~\AA\  at simple one parameter uncertainty - based  confidence level of 3.7~$\sigma$, consistently with several earlier LETG/ACIS reports.
Given the high statistical quality of the RGS1 data, the possibility of RGS1 accidentally missing the true line at $\lambda~\sim$~20~\AA\
is very low,  0.006\%. Neglecting this, the  LETG/ACIS detection can be interpreted as Ly$\alpha$ transition of \ion{O}{VIII} at one of the redshifts (z$\approx$~0.054) of FUV-detected warm WHIM. Given the very convincing X-ray spectral evidence for and against the existence of the $\lambda \sim$~20~\AA\ feature, we cannot conclude whether or not it is a true astrophysical absorption line.
Considering cosmological simulations,
the probability of LETG/ACIS $\lambda~\sim~20$~\AA\ feature being due to astrophysical \ion{O}{VIII} absorber co-located with the FUV-detected \ion{O}{VI} absorber is at the very low level level of $\lesssim$~0.1\%. We cannot rule out completely the very unlikely possibility that the  LETG/ACIS 20~\AA\ feature is due to a transient event located close to the blazar. 
}


\titlerunning{Tracing the hot WHIM in the \PKS\ sight line}

\authorrunning{J. Nevalainen et al.}

   \maketitle

\section{Introduction}
\label{intro}
High resolution X-ray spectroscopy is currently a popular method for searching for the local (z$<$1) missing baryons.
According to cosmological large scale simulations \citep[e.g.][]{1999ApJ...514....1C, 2001ApJ...552..473D, 2006MNRAS.370..656D,2009ApJ...697..328B,2012MNRAS.423.2279C, 2018MNRAS.473...68C} these missing baryons reside in the hottest (T~$\gtrsim10^{5.5}$~K) phase of the Warm Hot Intergalactic Medium (WHIM), embedded within the filaments of the Cosmic Web.
Given a bright enough background X-ray emission source, and a high enough column density of the intervening hot WHIM filament,  detectable high ion metal absorption line features (e.g. \ion{O}{VII}~He$\alpha$ and \ion{O}{VIII}~Ly$\alpha$) can be imprinted in the emission spectrum. 

Due to relatively weak X-ray signal of WHIM, compared to the sensitivities of the current instrumentation,
significant measurements (at statistical significance $>$~3~$\sigma$) of absorption lines from the ionised metals in the WHIM embedded within the large scale filaments are very sparse.
The rare 4~$\sigma$ detection of the \ion{O}{VII}~He$\alpha$ line at the Sculptor Wall  \citep{2009ApJ...695.1351B, 2010ApJ...714.1715F}
  has been questioned by 1) the existence of a galaxy 240 kpc away from the absorber \citep{2013ApJ...762L..10W} and 2) the possible contamination by the Galactic \ion{O}{II}~K$\beta$ absorption
  \citep{2016MNRAS.458L.123N} whose rest wavelength $\lambda~\approx$~22.30~\AA\ matches that of \ion{O}{VII}~He$\alpha$ at z=0.03, as in the case of the Sculptor Wall.
  We showed in  \citet{2015A&A...583A.142N} that an unrealistically long path length through the halo of the nearby galaxy is required for the hydrogen column density inferred from the X-ray measurement in Sculptor Wall. Also, the X-ray measurements of the Galactic \ion{O}{II}~K$\beta$ are questionable since the modeling heavily relies on the \ion{O}{II}~K$\alpha$ line whose
  wavelength $\lambda~\approx$~23.35~\AA\ co-incides with the poorly calibrated instrumental feature in both RGS and LETG, resulting in significantly different equivalent widths depending on which instrument is used (see \citet{2017A&A...605A..47N} and the references therein). Furthermore, the spectral modelling
  in \citet{2016MNRAS.458L.123N}, if assumed correct,  does not rule out a significant amount of \ion{O}{VII}~He$\alpha$ absorption at z=0.03 ($\log{N(\ion{O}{VII}({\rm cm}^{-2}))}~=$~15.4).

 \citet{2018Natur.558..406N} reported on the search for the WHIM absorption line in the sight line towards a blazar 1ES~1553+113.
When considering the redshift trials for the blind search and the systematic uncertainties of the instrument, they determined an absorption line - like feature at 
$\lambda~\sim~30.98$~\AA\ at a 3.5~$\sigma$ confidence level and interpreted it being due to \ion{O}{VII}~He$\alpha$ at z$\sim$0.43. It is not clear, whether the signal comes from WHIM located in a large scale filament
  (as required for the solution for the cosmological missing baryons problem) or from the halo of a nearby galaxy (as the authors prefer). Thus, the interpretation of the signal as due to \ion{O}{VII} within a large scale filament needs to be confirmed by detection of a galaxy filament using sufficiently deep spectroscopic data. The significance of the second line interpreted as 
\ion{O}{VII} at z$\sim$0.36 discussed in the paper has a significance below the 3~$\sigma$ limit.

On the other hand, the commonly used background sources, the blazars, are much brighter in the FUV compared to X-rays.
Also, the current FUV instruments (COS and STIS on-board HST; FUSE) are more sensitive than the current high resolution X-ray instruments (XMM-Newton/RGS and Chandra/LETG). Thus, the FUV measurements have yielded 
numerous detections of the extragalactic \ion{O}{VI} and broad~Ly$\alpha$ (BLA) absorption lines, typically interpreted as signatures of the warm (T~$\lesssim 10^{5.5}$~K) WHIM \citep[e.g.][]{2007ApJ...658..680L}. Thus, the warm part of the local WHIM is considered to be robustly detected \citep{2012ApJ...759...23S}.

Given the abundance of the FUV absorbers, it is tempting to use their locations to look for the hottest WHIM.
The underlying assumption of the co-location of the warm (\ion{O}{VI} and BLA) and hot (\ion{O}{VII-VIII}) WHIM absorbers has not been well tested yet, basically due to the very limited number of significant \ion{O}{VII-VIII} detections.
We utilised the above assumption by performing an X-ray follow-up in the sight line of a blazar  \PKS\ at the
redshifts where previous FUV measurements of \ion{O}{VI}, \ion{Si}{IV} and BLA absorption have indicated the existence of the warm WHIM.

\citet{2009ApJ...697.1784Y} stacked the available LETG/ACIS and MEG/ACIS data of \PKS\ together with the spectra from several other AGN sight lines
in order to examine the combined signal of the possible hot counterparts to the FUV-detected \ion{O}{VI} absorbers. They obtained no significant absorption line and upper limits of 
$\log{N(\ion{O}{VII}({\rm cm}^{-2}))}~\le~14.6$ and $\log{N(\ion{O}{VIII}({\rm cm}^{-2}))}~\le~15.5$.
Compared to \citet{2009ApJ...697.1784Y} we have the luxury of a very large amount of high resolution X-ray data on \PKS .
We used all useful data available to us on \PKS\ obtained with RGS1 on-board XMM-Newton (the RGS2 does not cover most of the studied lines) and
LETG/ACIS and LETG/HRC-S combinations on-board Chandra (the quality of the available MEG/ACIS data was too poor to yield meaningful constraints).
 We utilised these data in order to obtain similar detection limits as \citet{2009ApJ...697.1784Y} but for the individual FUV 
absorbers in a single sight line (i.e. \PKS ), as reported by \citet{2012ApJ...759..112T}.
While doing so, we will avoid possible problems due to stacking the data from different instruments, targets and redshifts.
With this we aim at improving the observational status of the possible co-location of the warm and hot WHIM.

We use $\Omega_{\rm m} = 0.3$, $\Omega_{\Lambda} = 0.7$ and $H_0 = 70~\mathrm{km~s}^{-1}\mathrm{Mpc}^{-1}$.
The distances and redshifts refer to the heliocentric frame. We quote uncertainties at the 1~$\sigma$ confidence level.

\begin{table*}[t]
\centering
\caption{The FUV WHIM absorbers in the sight line to blazar \PKS .}
\small{
\label{tab:absorbers}
\begin{tabular}{lcccccccccc}
\hline\hline

ID\tablefootmark{a}    &   $z_{FUV}$\tablefootmark{b} &  \multicolumn{6}{c}{line}        & $z_{X-ray}$\tablefootmark{c}  & $\lambda$(\ion{O}{VII}~He$\alpha$)\tablefootmark{d}  & $\lambda$(\ion{O}{VIII}~Ly$\alpha$)\tablefootmark{d}  \\
                       &         & \multicolumn{3}{c}{\bf Metal}  &   \multicolumn{3}{c}{\bf BLA}           &                                &                         &                    \\
                       &         & ion & EW\tablefootmark{e} & $\log{N({\rm cm}^{-2})}$\tablefootmark{f}  & EW\tablefootmark{g}  &  b\tablefootmark{h}   & $\log{N({\rm cm}^{-2})}$\tablefootmark{i}     &     & &                           \\
                       &         &     & (m\AA)  &   & (m\AA)   &  (km s$^{-1}$)   &     &   & (\AA)  &       (\AA)                      \\
\hline
 A1                    & 0.00878 & -- & -- &         & 49$\pm$8    &  59$\pm$8   & 13.0$\pm$0.1         &   0.00878    & 21.79 &  19.14    \\
 A2                    & 0.01892 & -- & -- &         & 59$\pm$4    &  38$\pm$4   &  13.0$\pm$0.1          &   0.01892    & 22.01 &  19.33     \\
 A3a\tablefootmark{j}  & 0.05405 & \ion{O}{VI} & 32$\pm$5\tablefootmark{k}, 30$\pm$9\tablefootmark{k} & 13.6$\pm0.1$  & 315$\pm$4 &  44$\pm$0 & 14.1$\pm$0.1 & \multirow{2}{*}{0.05425}    & \multirow{2}{*}{22.77} &  \multirow{2}{*}{20.00}    \\
 A3b\tablefootmark{j}  & 0.05445 & -- & -- &         & 60$\pm$31   &  54$\pm$7   & 13.0$\pm$0.1          &              & &        \\
 A4a\tablefootmark{j}  & 0.05659 & -- & -- &         & 477$\pm$10  &  48$\pm$1   & 14.5$\pm$0.3          & \multirow{2}{*}{0.05683}    & \multirow{2}{*}{22.83} &  \multirow{2}{*}{20.05}    \\
 A4b\tablefootmark{j}  & 0.05707 & \ion{O}{VI}  & 44$\pm$11\tablefootmark{l} & 13.6$\pm$0.1 & 424$\pm$11   &  68$\pm$1   & 14.0$\pm$0.0        &                            & &      \\
 A5                    & 0.08062 & \ion{Si}{IV} & 12$\pm$4\tablefootmark{m}  & 12.1$\pm$0.1        & 29$\pm$5    &  40$\pm$5     & 12.7$\pm$0.1       &   0.08062    & (23.34)\tablefootmark{n} & 20.50    \\
 A6a\tablefootmark{j}  & 0.10552 & --  & -- &        & 360$\pm$7   &  54$\pm$1   &  14.1$\pm$0.2        &  \multirow{2}{*}{0.10569}  & \multirow{2}{*}{23.83} &  \multirow{2}{*}{20.97}       \\
 A6b\tablefootmark{j}  & 0.10586 & --  & -- &        & 156$\pm$22  &  66$\pm$4   &  13.5$\pm$0.0        &      & &    \\
 \hline
\end{tabular}
\tablefoot{
\tablefoottext{a}{Name of the absorber}
\tablefoottext{b}{The redshift of the FUV absorber.}
\tablefoottext{c}{Our adopted redshift for the X-ray follow-up.}
\tablefoottext{d}{The redshifted wavelength of \ion{O}{VII}~He$\alpha$ and \ion{O}{VIII}~Ly$\alpha$.}
\tablefoottext{e}{The equivalent width of the metal line.}
\tablefoottext{f}{The column density of the metal line.}
\tablefoottext{g}{The equivalent width of the BLA.}
\tablefoottext{h}{The Doppler parameter of the BLA.}
\tablefoottext{i}{The column density of the BLA.}
\tablefoottext{j}{The two FUV absorbers at the same WHIM structure candidate are shown separately.}
\tablefoottext{k}{The equivalent width of the \ion{O}{VI} 1031.9\AA\ and 1037.6\AA\ transitions, respectively.}
\tablefoottext{l}{The equivalent width of the only significantly detected \ion{O}{VI} 1031.9\AA\ transition.}
\tablefoottext{m}{The equivalent width of the only significantly detected \ion{Si}{IV} 1393.8\AA\ transition.}
\tablefoottext{n}{The possible A5  \ion{O}{VIII} line would land at problematic wavelengths due to astrophysical and instrumental oxygen edges which prevents an accurate modelling. We thus excluded this line from the further analysis.}
}
}
\end{table*}

\section{FUV detections}
\label{FUV}
In the present work we employed the catalogue of blazar FUV measurements from \citet{2012ApJ...759..112T}.
The \PKS\ sight line has been extensively studied using the Hubble Space Telescope COS instrument
\citep{2014ApJS..212....8S} and  \citep{2016ApJ...832...76D} and the Far Ultraviolet Spectroscopic Explorer (FUSE) and Space Telescope Imaging Spectrograph (STIS) 
(e.g. \citet{1998AJ....116.2094S, 2003ApJ...594L.107S} ;
\citet{2003ApJS..146..165S} ;
\citet{2003ApJS..146....1W} ;
\citet{2008ApJ...679..194D} ;
\citet{2013ApJ...763..148S, 2014ApJ...791..128S}
and \citet{2017A&A...607A..48R}).
The inclusion of the FUSE and the STIS instruments ensures the coverage of \ion{O}{VI} lines at all redshifts of interest due to their broader wavelength range, compared to the COS.
 
The BLA and \ion{O}{VI} absorption, observable in the FUV band, are commonly interpreted as originating from the warm WHIM. Using a single \ion{O}{VI} transition of the 1031.9/1037.6~\AA\ doublet or a single BLA as a WHIM signature may be too optimistic, given the possibilities of misidentification of a single line.  
On the other hand, a more robust criterion of requiring both \ion{O}{VI} transitions to be significantly detected reduces the number of the potential 
WHIM redshifts due to e.g. detector gap at the critical wavelengths, and consequently we may loose some true WHIM signal. Thus, we adopted the more relaxed criterion of requiring at least one significantly (3$\sigma$) detected \ion{O}{VI} line (or other metal line whose ionisation temperature exceeds $10^5$ K)
or BLA for follow-up, keeping the above caveat in mind.

As BLA we considered absorption lines due to \ion{H}{I}~Ly$\alpha$ transition with broadening velocity higher than 40~km~s$^{-1}$, which corresponds to thermal broadening of hydrogen at T = $10^5$ K, the lower limit of the WHIM temperature range.

Using the above criteria, the catalogue of \citet{2012ApJ...759..112T} yielded \ion{O}{VI}, \ion{Si}{IV} or BLA detections at 9 different redshifts
in the \PKS\ sight line (see Table~\ref{tab:absorbers}). 
Considering the LETG/RGS photon energy resolution of 40-60~m\AA\ at $\lambda~\approx$ 20~\AA , we associated the FUV absorbers with redshift difference smaller than 0.002 (co-moving distance difference of $\sim$8 Mpc) with a single X-ray absorber. For such absorbers (i.e. A3, A4 and A6) we use the average redshift of the FUV lines 
as the X-ray follow-up redshift. As a result we have the redshifts of six possible X-ray absorbers to study with X-ray instruments in the \PKS\ sight line (see Table~\ref{tab:absorbers}).

\section{X-ray analysis}
\label{xray}
We examine here the most likely hot WHIM tracers, i.e. \ion{O}{VII} He$\alpha$  and \ion{O}{VIII} Ly$\alpha$ lines at the five FUV-predicted redshifts
(see Table~\ref{tab:absorbers}), including the much studied and controversial \ion{O}{VIII} Ly$\alpha$ line at z=0.054-0.056 (see Section~\ref{history}).

\begin{table*}
 \centering
  \caption{The equivalent widths and the column densities for the hot WHIM.}
 \label{results.tab}
 \begin{tabular}{lcccc}
 \hline\hline
 instr. & \multicolumn{2}{c}{\ion{O}{VII} He$\alpha$}  & \multicolumn{2}{c}{\ion{O}{VIII} Ly$\alpha$} \\
        & EW\tablefootmark{a}  & $\log{N(ion)}$\tablefootmark{a} & EW\tablefootmark{a}  & $\log{N(ion)}$\tablefootmark{a} \\
        & m\AA                 & cm$^{-2}$                        & m\AA                 & cm$^{-2}$             \\
 \hline
\smallskip \\
\multicolumn{5}{c}{\bf A1 (z = 0.00878 ; $\lambda_{\ion{O}{VII}}$ = 21.79~\AA   ;   $\lambda_{\ion{O}{VIII}}$ = 19.14~\AA)} \\
RGS1   & 2.2$\pm$1.8 & 14.9$^{+0.3}_{-0.8}$  & $<$2.1 & $<$15.2 \\  
HRC    & $<$2.6 & $<$15.0  & $<$1.2 & $<$15.0 \\
\smallskip \\
\multicolumn{5}{c}{\bf A2 (z = 0.01892 ; $\lambda_{\ion{O}{VII}}$ = 22.01~\AA   ;   $\lambda_{\ion{O}{VIII}}$ = 19.33~\AA)} \\
RGS1  & $<$1.2 & $<$14.6 & $<$2.0 & $<$15.2 \\  
HRC   & $<$1.0 & $<$14.6 & $<$2.4 & $<$15.3 \\
\smallskip \\
\multicolumn{5}{c}{\bf A3 (z = 0.05425 ; $\lambda_{\ion{O}{VII}}$ = 22.77~\AA   ;   $\lambda_{\ion{O}{VIII}}$ = 20.00~\AA)} \\
RGS1  & $<$4.0 & $<$15.2 & $<$1.0 & $<$14.9 \\  
HRC   & $<$4.6 & $<$15.2 & $<$2.3 & $<$15.2 \\
\smallskip \\
\multicolumn{5}{c}{\bf A4 (z = 0.05683 ; $\lambda_{\ion{O}{VII}}$ = 22.83~\AA   ;   $\lambda_{\ion{O}{VIII}}$ = 20.05~\AA)} \\
RGS1  & $<$0.8 & $<$14.5 & $<$1.4 & $<$15.0 \\  
HRC   & $<$5.8 & $<$15.4 & $<$2.3 & $<$15.3 \\
\smallskip \\
\multicolumn{5}{c}{\bf A5 (z = 0.08062 ; $\lambda_{\ion{O}{VII}}$ = 23.34~\AA   ;   $\lambda_{\ion{O}{VIII}}$ = 20.50~\AA)} \\
RGS1  & --\tablefootmark{b} & --\tablefootmark{b} & $<$1.9 & $<$15.2 \\  
HRC   & --\tablefootmark{b} & --\tablefootmark{b} &  $<$2.6 & $<$15.3  \\
\smallskip \\
\multicolumn{5}{c}{\bf A6 (z=0.10569  ;  $\lambda_{\ion{O}{VII}}$ = 23.89~\AA   ;   $\lambda_{\ion{O}{VIII}}$ = 20.97~\AA)} \\
RGS1  & $<$1.8 & $<$14.8 & $<$1.4 & $<$15.0 \\  
HRC   & $<$1.6 & $<$14.8 & $<$2.6 & $<$15.3 \\
\hline
\end{tabular}
\tablefoot{
\tablefoottext{a}{The constrains or the upper limits at 1$\sigma$ confidence level for the equivalent width (EW) and the ion column densities $\log{N(ion)}$, where the uncertainties include both the statistical one and the systematic 2\% of the flux.}
\tablefoottext{b}{The wavelengths coincide with the instrumental feature}
}
\end{table*}

\subsection{X-ray data}

We analysed all \PKS\ spectra obtained with RGS1 and LETG/HRC-S available in the year 2016, published in \citet{2017A&A...605A..47N}.
The RGS2 does not cover most of the interesting wavelengths and we thus ignore that data. We examine LETG/ACIS spectra separately in Section~\ref{acis}. 
The exposure time of the RGS1 exceeds 1 Ms rendering the data very powerful for measuring the weak WHIM lines.

We analysed the data using SPEX spectral fitting package  \citep{1996uxsa.conf..411K}.  
We used 20 m\AA\ bin size for RGS and 25 m\AA\ for LETG/HRC, oversampling the spectral resolution by a factor of 2-3.
Due to the high number of counts in the spectral channels we used $\chi^{2}$ statistics.
We added 2\% of the flux at each channel as a systematic uncertainty of the effective area in quadrature to the statistical uncertainties, 
see \citet{2017A&A...605A..47N}.

\subsection{Wavelength scale calibration}
\label{gallines}
To correct for possible wavelength scale calibration offsets, the individual RGS spectra have been shifted as indicated by the Galactic neutral absorber lines before co-addition
(see the description and references in \citet{2017A&A...605A..47N}).

The same procedure was not practical in the case of HRC spectra, since its lower effective area and exposure time ($\sim$300 ks) compared to RGS, rendered the statistical quality too poor for measuring the Galactic line centroids accurately using the individual spectra. Thus, we examined the total HRC spectrum, co-added without any shifts. 
The Galactic \ion{O}{I} ($\lambda$ = 23.51 \AA ) and \ion{O}{VII} ($\lambda$= 21.60 \AA) lines were unambiguously identified.
We fitted the 23.0--24.0 \AA\ and 21.0--22.0 \AA\ bands with a model consisting of a power-law continuum and a Gaussian line and consequently obtained significant detections of the Galactic \ion{O}{I} and \ion{O}{VII} lines. Their centroid wavelengths were detected with an accuracy better than 
$\pm$~10~m\AA . Within the uncertainties, the centroid wavelengths agreed with the a priori values. Thus, the HRC energy scale of the co-added spectrum was very accurately calibrated.

\subsection{Wavebands}
Our aim is to cover the possible \ion{O}{VII}~He$\alpha$ and \ion{O}{VIII}~Ly$\alpha$ lines, assuming the redshifts of the absorbers A1-A6
(see Table~\ref{tab:absorbers} for the wavelengths). 
We wish to select minimally wide bands around the above lines to simplify the continuum modelling, but wide enough to obtain a robust continuum level.
We decided to confine the analyses within wavebands 21.7--24.0~\AA\ (the \ion{O}{VII} band) and 19.05--21.1 \AA\ (the \ion{O}{VIII} band). 

We excluded some problematic channels, as described in the following.
In both RGS1 and HRC the wavelength of the A5~\ion{O}{VII} absorber ($\lambda \approx$ 23.34~\AA ) coincides with the instrumental absorption feature 
\citep{2003A&A...404..959D,2015A&A...573A.128D} rendering the measurement inaccurate. We thus excluded the 22.85--23.6~\AA\ band from both spectra and consequently we will not analyse the possible A5~\ion{O}{VII}. Due to the RGS1 CCD4-5 gap we excluded the 20.75--20.90 \AA\ band from the RGS1 analysis. Fortunately none of our candidate lines has the centroid in that band. There is excess of HRC data on top the otherwise smooth continuum at 19.74--19.85~\AA\  which cannot be adequately fitted. 
We thus excluded the HRC data in this band which does not contain any of the tested lines.

\subsection{Galactic absorption}
\label{galabs}
For the ionised Galactic absorption, we adopted the \citet{2017A&A...605A..47N} model: the hot halo (HH) and the transition temperature gas (TTG), 
manifested as \ion{O}{IV}, \ion{O}{V}, \ion{O}{VII}, \ion{O}{VIII}, \ion{N}{VI},\ion{C}{VI} and \ion{Ne}{IX} absorption lines, were modelled
as two CIE absorbers (a model called hot in SPEX). We allowed the parameters of these components to vary within the statistical 1~$\sigma$ uncertainties
as derived from RGS1 data in \citet{2017A&A...605A..47N}:\\
N(H$_{HH}$) = 2.4$\pm$0.3 $\times 10^{19}$ cm$^{-2}$,
kT$_{HH}$ = 1.7$\pm$0.1 $\times 10^{-1}$ keV,
N(H$_{TTG}$) = 1.0$\pm$0.2 $\times 10^{19}$ cm$^{-2}$,
kT$_{TTG}$ = 1.4$\pm$0.1 $\times 10^{-2}$ keV.
We assumed the element number density ratios from \citet{2009M&PSA..72.5154L} and that the metal abundance is Solar.
We constrained the non-thermal broadening velocity to 15--35~km~s$^{-1}$, as in \citep{2017A&A...605A..47N} based on the FUV measurements of Wakker (priv. comm.).

For the LETG/HRC, we additionally included the neutral Galactic disc absorber consisting of atomic and molecular components
which have been used to correct for the cold Galactic absorption when processing the RGS spectra (see \cite{2017A&A...605A..47N} for details).

The band passes occupied by the possible \ion{O}{VII} line due to A3 and A4 overlap with that of the Galactic \ion{O}{IV} line (see \citet{2017A&A...605A..47N} for the discussion of the identification of the line as Galactic \ion{O}{IV}).
However, since we used priors in the Galactic absorber model (see above), the fit converged to a $\chi^2$ minimum.

\subsection{Blazar emission}
The RGS spectra have already been normalised to the \PKS\ continuum, i.e. the \PKS\ emission absorbed by the cold Galactic component, see \citet{2017A&A...605A..47N}.
Thus, we modelled here the RGS blazar emission with a constant, which we allowed to vary in order to accommodate for the statistical uncertainties of 
the continuum modelling.

The emission spectrum of \PKS\ as measured with HRC was adequately modelled in the \ion{O}{VII} band with a power-law. 
In the WHIM analysis below, we allowed the normalisation and the photon index of the power-law component to vary.
However, in the case of the HRC \ion{O}{VIII} band 
the continuum was more complicated.
We modelled that with a spline model with a constant grid size of 300 m\AA.
Allowing all the spline parameters to be free in the WHIM analysis allowed too much freedom to the model in the sense that some of the tested lines
were significantly affected.
Thus we fixed the spline parameters to their best fit values in order to maintain the continuum shape and allowed only the normalisation of the continuum to vary. 

\subsection{WHIM lines}
In order to examine the possible WHIM lines, we added a SPEX model called slab to the continuum model described above.
``Slab'' calculates the transmission of a thin slab of material whose ion column densities can be varied independently, i.e. the ion ratios are not determined by the ionisation balance.
The Lorentz component of the final Voigt profile is calculated for each transition in the SPEX atomic data base for a given ion,
while the Gaussian component is calculated based on the input value of the total velocity dispersion (thermal and non-thermal).
We fixed the total broadening to 100~km~s$^{-1}$, which corresponds to the pure thermal broadening of oxygen at T = 10$^7$~K.
Thus, in the case of significant non-thermal broadening, our measurements of the column densities of \ion{O}{VII} and \ion{O}{VIII} are somewhat overestimated. We get back to this point when discussing the measurements below.

We fitted the data of each instrument separately, fixing the slab model wavelength to that of either \ion{O}{VII}~He$\alpha$ or \ion{O}{VIII}~Ly$\alpha$, 
applying one of the adopted X-ray redshifts at a time for a given fit (see Table~\ref{tab:absorbers}). The column density of a given ion was the only free parameter.

We obtained the best-fit values and the uncertainties (including both the statistical and the systematic 2\% of the flux)  
of the column densities by the $\chi^2$ minimisation. 
We used these models to calculate the equivalent width and its uncertainties (EW$\pm \sigma_{EW}$) of each tested line 
and used the ratio EW / $\sigma_{EW}$ as a measure of its detection significance.

\subsection{Results}
\begin{figure*}
\includegraphics[width=18cm,angle=0]{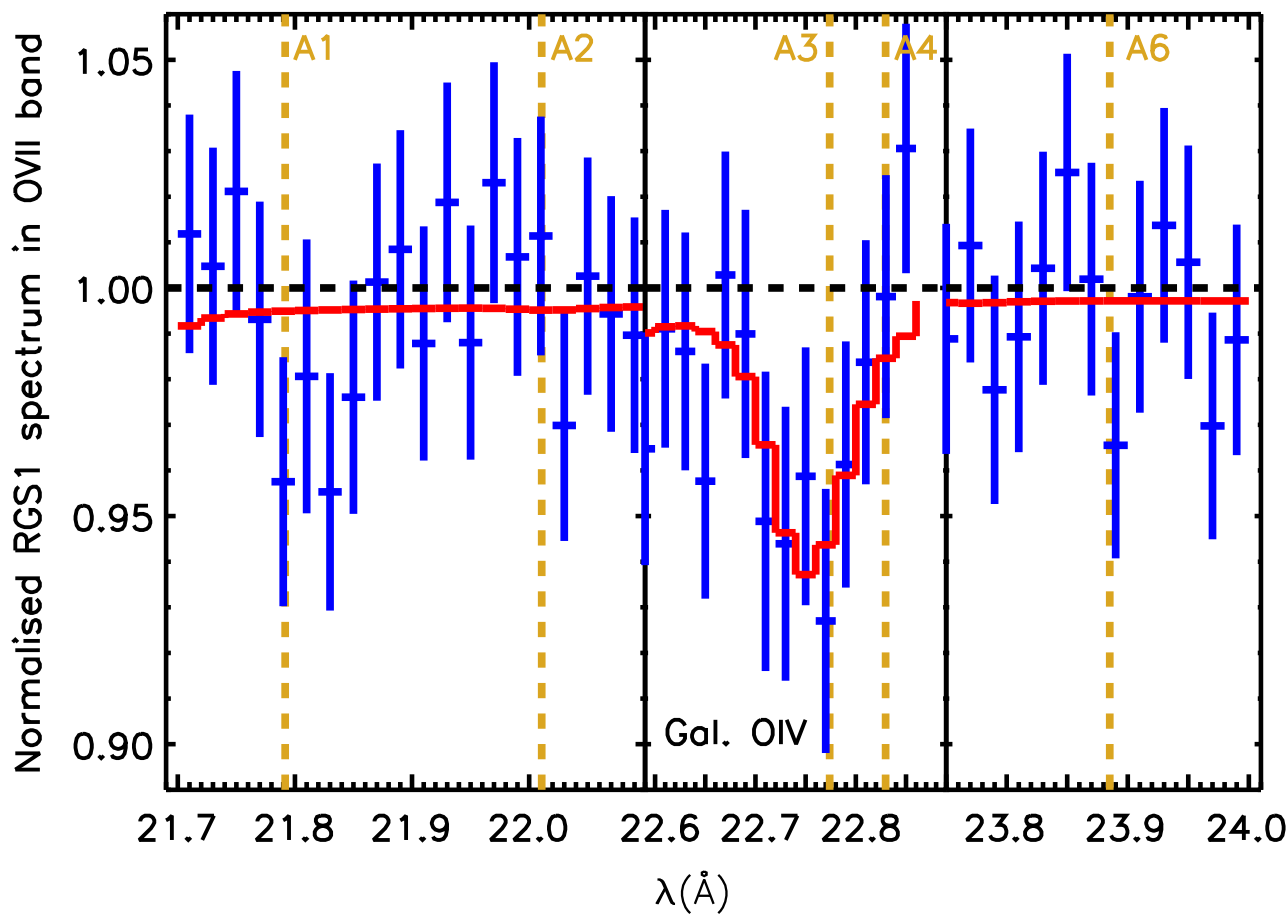}
\caption{The normalised RGS1 data (blue crosses) and the normalised best-fit model (red line) for the \PKS\ emission absorbed with the Galactic neutral disk, hot halo and the transition temperature gas TTG (see \citet{2017A&A...605A..47N}). The normalisation is done by dividing the spectra by the best-fit model consisting of the \PKS\ emission and only the Galactic neutral absorber to high-light the effect of the Galactic halo and TTG. In particular, the Galactic \ion{O}{IV} absorption line overlaps with the \ion{O}{VII} test line for A3 (middle panel). The dashed vertical lines indicate the centroid wavelengths of the \ion{O}{VII} test lines assuming the redshifts of the structures A1, A2, A3, A4 and A6 (A5 is missing due to the overlap of its centroid wavelength with the instrumental feature in both RGS1 and LETG, see \citet{2003A&A...404..959D,2015A&A...573A.128D}). The error bars contain both the statistical uncertainties and the systematic one (2\% of the flux).}
\label{WHIM_OVII_RGS1.fig}
\end{figure*}

\begin{figure*}
\includegraphics[width=18cm,angle=0]{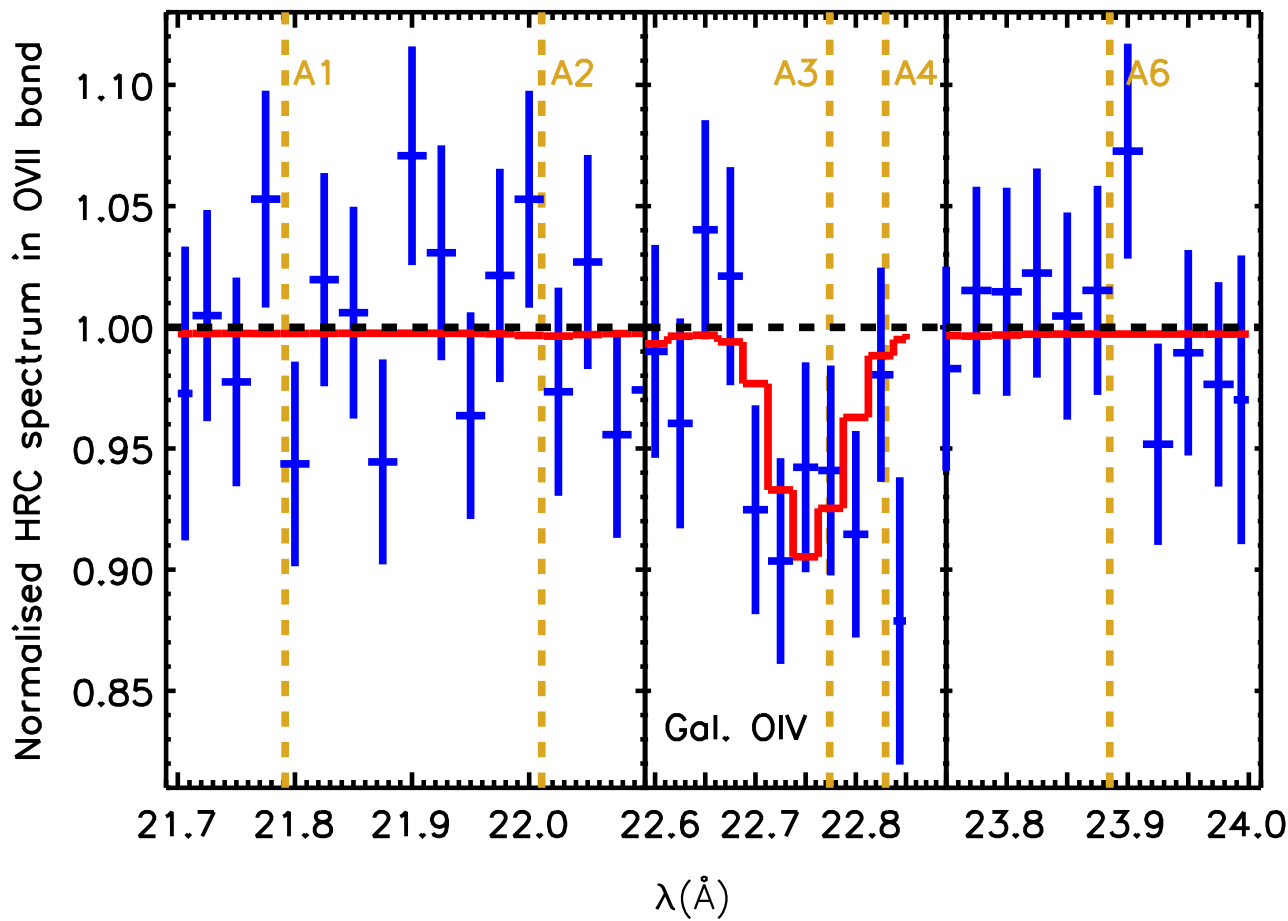}
\caption{As Fig.~\ref{WHIM_OVII_RGS1.fig} but for the LETG/HRC instrument combination. 
}
\label{WHIM_OVII_HRC.fig}
\end{figure*}

\begin{figure*}
\includegraphics[width=18cm,angle=0]{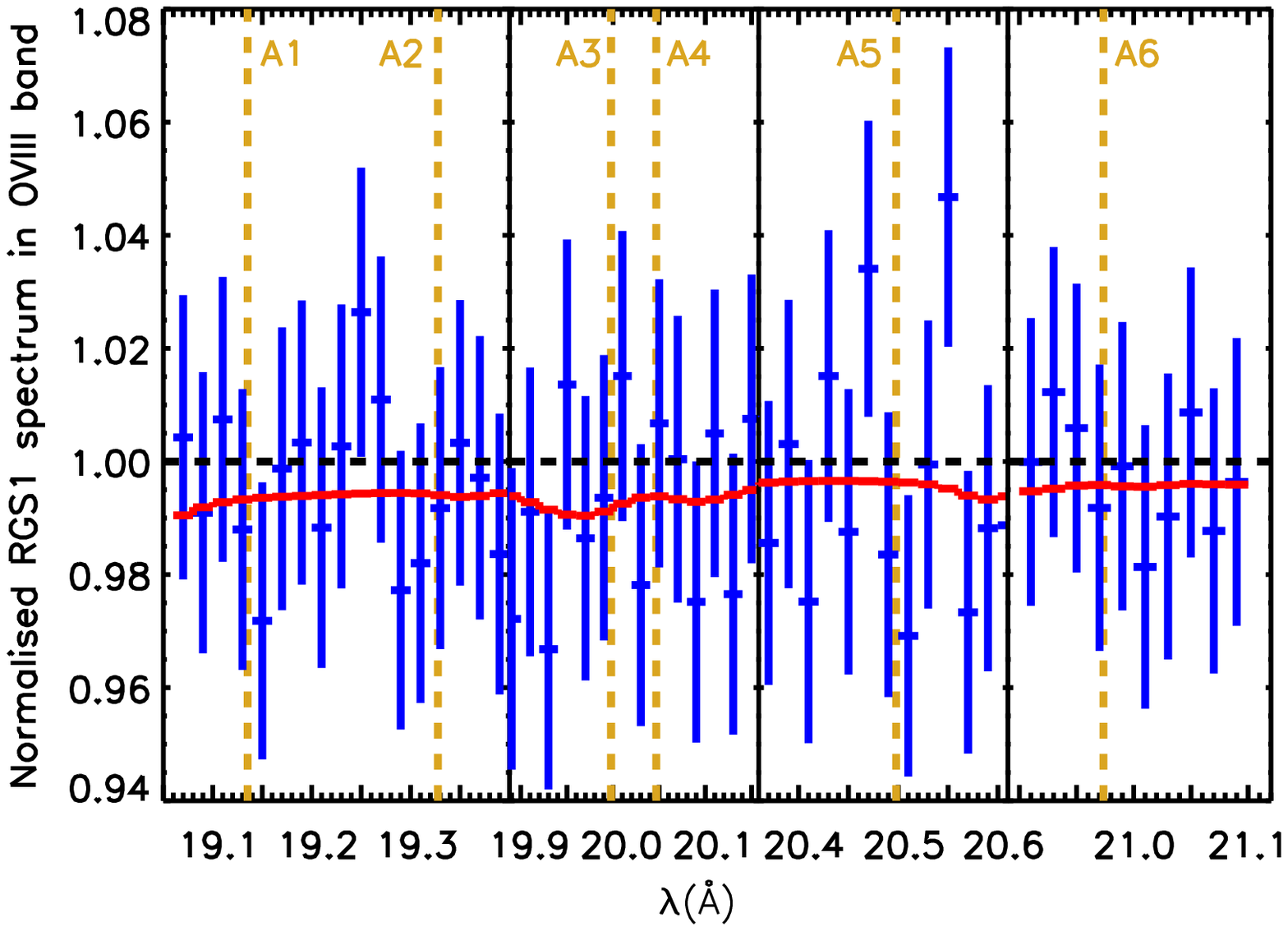}
\caption{As Fig.~\ref{WHIM_OVII_RGS1.fig} but for the \ion{O}{VIII} line.}
\label{WHIM_OVIII_RGS1.fig}
\end{figure*}

\begin{figure*}
\includegraphics[width=18cm,angle=0]{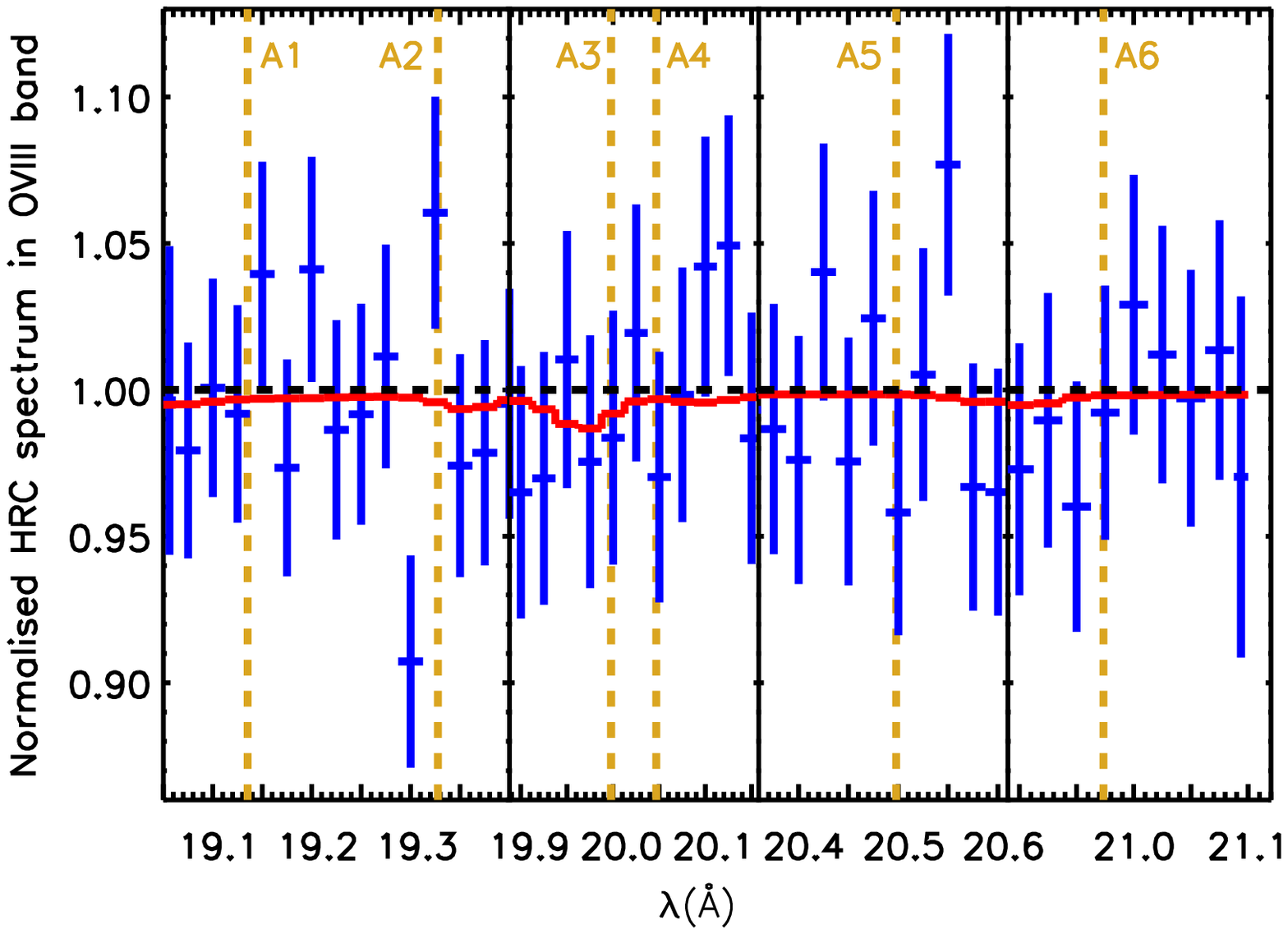}
\caption{As Fig.~\ref{WHIM_OVII_HRC.fig} but for the \ion{O}{VIII} line.}
\label{WHIM_OVIII_HRC.fig}
\end{figure*}

We found that none of the tested lines was detected significantly with RGS1 or HRC
(see Figs.~\ref{WHIM_OVII_RGS1.fig}~-~\ref{WHIM_OVIII_HRC.fig} and Table~\ref{results.tab}).
The upper limits\footnote{The obtained upper limits for the column densities may be slightly overestimated due to neglected, possible non-thermal velocities.}
of a few m\AA\ correspond to \ion{O}{VII} and \ion{O}{VIII} level of $\sim 10^{15}$ cm$^{-2}$.
We discuss the implications of these results in Section~\ref{impli}.
 
The underlying assumption in the above analysis was that the warm and hot WHIM are co-located and at rest (or moving with the same velocity), so that z$_{X-ray}$ = z$_{FUV}$.
We next relaxed this assumption by allowing a relative projected sight line velocity difference up to $\pm$600~km~s$^{-1}$
(i.e. $\Delta_Z$ =  $\pm$0.002)
between the FUV and X-ray absorbers,
while co-located. In practise, we repeated the above fits, but shifting the test line centroid by $\pm$40 m\AA\ in steps of 20 m\AA . 
This resulted in no significant change in the results\footnote{We found marginal ($<$2 $\sigma$) indication for \ion{O}{VII} absorption at z = 0.0101$\pm$0.0006 in the RGS1 data. 
The redshift change corresponds to velocity difference of $\sim$400~km~s$^{-1}$ compared to the FUV-based A1 X-ray test redshift.}.

\begin{figure*}
\includegraphics[width=18cm,angle=0]{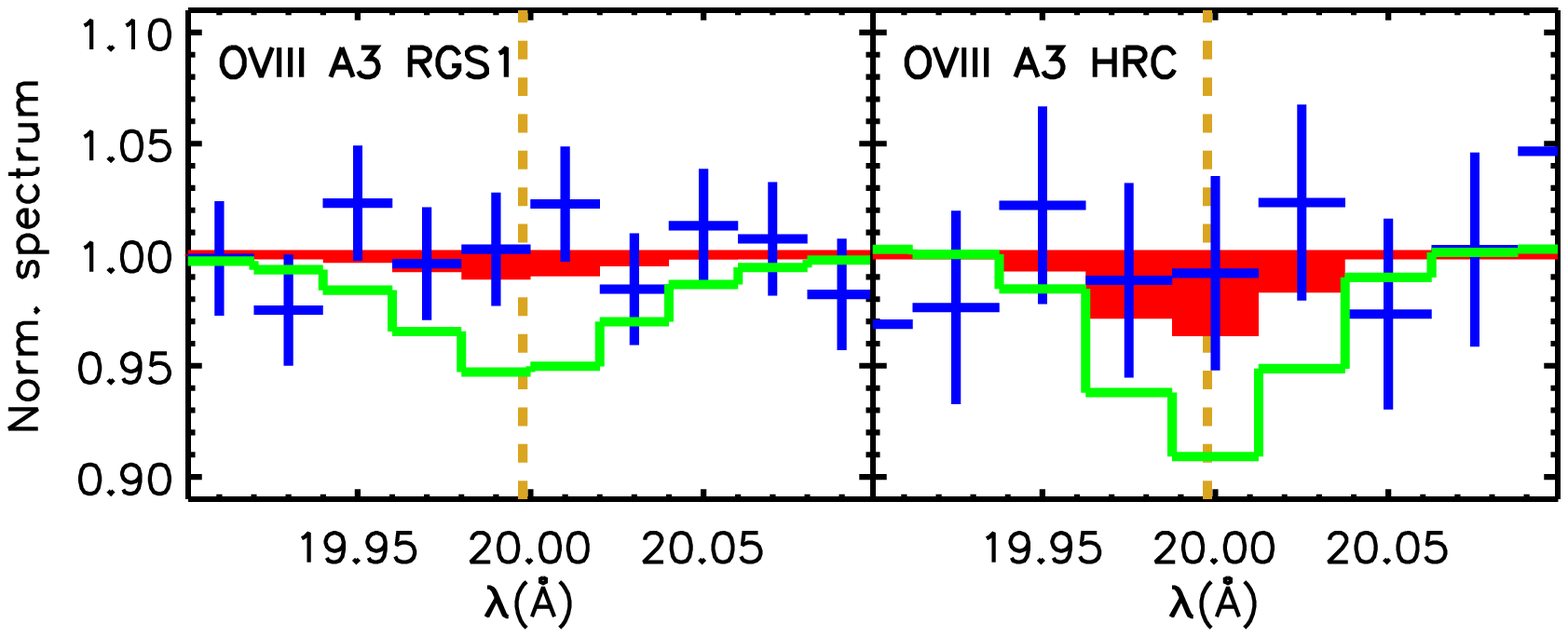}
\vspace{-5cm}
\caption{The normalised data (blue crosses) and the \ion{O}{VIII} line models at the upper 1$\sigma$ limit (red areas) 
at the redshift of A3 for RGS1 (left panel) and LETG/HRC-S (right panel). 
The normalisation is done by dividing the spectra by the best-fit model consisting of the \PKS\ emission and all (neutral + ionised) Galactic absorbers.
The vertical dashed lines indicate the centroid wavelength of the \ion{O}{VIII} doublet redshifted to z=0.0544.
The green lines indicate the lower limit reported by \citet{2007ApJ...670..992F}. The error bars contain both the statistical uncertainties and the systematic one (2\% of the flux).}
\label{WHIM_OVIII_A3_noacis.fig}
\end{figure*}

\section{\ion{O}{VIII} Ly$\alpha$ absorption at z$\sim$0.054?}
Before discussing the implications of the above results, we investigate here the widely reported line-like feature at $\lambda \sim$20 \AA\ in the LETG/ACIS \PKS\ spectrum, interpreted as \ion{O}{VIII} Ly$\alpha$ absorption at z=0.054-0.056
 (see Table~\ref{OVIII_meas.tab}).
The redshift of such line agrees with the range of values of the FUV lines associated with our A3 absorber (see Table~\ref{tab:absorbers}).
We use a notation \ion{O}{VIII}$_{\rm A3}$ for this line in the following. 

\subsection{Summary of earlier X-ray work}
\label{history}

\begin{table*}
 \centering
    \caption{\ion{O}{VIII} Ly$\alpha$ measurements at z$\sim$0.054-0.056 in the \PKS\ sight line.}
 \label{OVIII_meas.tab}
  \begin{tabular}{lcccccccc}
  \hline\hline
                            &                &             &                &            &                &                   &                     &                \\
Ref.                        & \multicolumn{2}{c}{RGS1}     & \multicolumn{2}{c}{LETG/HRC}                 & \multicolumn{4}{c}{LETG/ACIS-S}                          \\
                            &                &             &                &            &                &                   &                     &                \\
                            & Exp. time      & EW\tablefootmark{f}          & Exp. time      & EW\tablefootmark{f}         & Exp. time      & z                 & EW\tablefootmark{f}                  & signif.\tablefootmark{g}        \\
                            & (ks)           & (m\AA )     & (ks)           & (m\AA )    & (ks)           &                   & (m\AA )             &                \\
F02\tablefootmark{a}        & --             & --          & --             & --         & 80              & 0.056$\pm$0.001   & 14.0$^{+4.4}_{-3.4}$  & 4.5 $\sigma$   \\                    
C04\tablefootmark{b}        & 110            & $\le$ 7     & --             & --         & --             & --                & --                  & --             \\ 
F07\tablefootmark{c}        & --             & --          & 200            & $\le$ 5    & 280            & 0.054$\pm$0.001   & 7.4$^{+1.7}_{-1.2}$   & 5.0 $\sigma$   \\
W07\tablefootmark{d}        & --             & --          & 230            & $\le$ 6    & 250            & 0.056$\pm$0.001   & 7.5$\pm{2.1}$       & 3.5 $\sigma$   \\
this work\tablefootmark{e}  & 1200           & $\le$ 1.3   & 310             & $\le$ 2.1  & 330            & 0.0554$\pm$0.0004 & 6.2$\pm{1.7}$       & 3.2 $\sigma$   \\ 
                            &                &             &                &            &                &                   &                     &                \\ 
\hline                  
\end{tabular}
\tablefoot{ 
\tablefoottext{a}{\citet{2002ApJ...572L.127F}}, 
\tablefoottext{b}{\citet{2004ApJ...603..449C}}, 
\tablefoottext{c}{\citet{2007ApJ...670..992F}}, 
\tablefoottext{d}{\citet{2007ApJ...665..247W},}
\tablefoottext{e}{The final sample,}
\tablefoottext{f}{The equivalent width is reported at 1~$\sigma$ confidence level,}
\tablefoottext{g}{The significance of the detection.}
}
\end{table*}

Combining three observations with total exposure time of 80 ks
obtained with LETG/ACIS-S, \citet{2002ApJ...572L.127F} reported a 4.5~$\sigma$ detection of an absorption line at $\lambda$ = 20.02$\pm$0.02 \AA\ and interpreted that as \ion{O}{VIII} Ly$\alpha$ absorption at z = 0.056$\pm$0.001 (see Table~\ref{OVIII_meas.tab} for the summary of different measurements).
This was contrasted by \citet{2004ApJ...603..449C} who studied 110 ks of RGS1 data of \PKS\ and did not detect the line,
despite of the longer exposure and higher effective area.
An another analysis with LETG/ACIS data \citep{2007ApJ...670..992F}, this time with a larger exposure time of 280~ks, compared to that in  \citet{2002ApJ...572L.127F},  yielded a 5.0~$\sigma$ detection of an absorption line at $\lambda$~=~20.00$\pm$0.01~\AA , consistent with the \ion{O}{VIII}~Ly$\alpha$ line at z~=~0.054$\pm$0.001. The equivalent width of the line is consistent with that of \citet{2002ApJ...572L.127F} value within the uncertainties at 90\% confidence level. 
\citet{2007ApJ...665..247W} analysed a LETG/ACIS data set, largely overlapping with \citet{2007ApJ...670..992F}, and detected an absorption line
at 3.5~$\sigma$ confidence level at  
$\lambda$~=~20.03$\pm$0.01~\AA, i.e.
at a bit higher wavelength (2~$\sigma$) than \citet{2007ApJ...670..992F}, corresponding to z~=~0.056$\pm$0.001 if interpreted as the \ion{O}{VIII}~Ly$\alpha$ line. The equivalent widths agree very well in the two works.
They also analysed 230~ks of LETG/HRC data, which however did not yield a significant detection.

\subsection{Our results}
We used our RGS1 non-detection of \ion{O}{VIII}$_{\rm A3}$ i.e. EW(\ion{O}{VIII}$_{\rm A3}$) $\le$ 1.3 m\AA\ (Table~\ref{results.tab}), to estimate the upper limit for the column density as $\log{N(\ion{O}{VIII}_{\rm A3}({\rm cm}^{-2}))} \le 14.9$.
This is by a factor of four smaller than the lower limit reported by \citet{2007ApJ...670..992F} and \citet{2007ApJ...665..247W} obtained with LETG/ACIS (Table~\ref{OVIII_meas.tab}).
Also our LETG/HRC analysis yielded only an upper limit of EW(\ion{O}{VIII}$_{\rm A3}$) $\le$ 2.1 m\AA, significantly lower than the 
reported LETG/ACIS detections (see Fig.~\ref{WHIM_OVIII_A3_noacis.fig} and Table~\ref{OVIII_meas.tab}).
We investigate in the following what might be causing such large discrepancies.

\subsection{RGS issues}
\label{RGSissues}
As discussed in Section~\ref{history}, the previous XMM-Newton/RGS work of a subsample of our data \citep{2004ApJ...603..449C} yielded no detection of 
\ion{O}{VIII}$_{\rm A3}$, while the upper limit for the equivalent width was inconsistent with the LETG/ACIS detections of
\citet{2007ApJ...670..992F} and \citet{2007ApJ...665..247W}.
Our RGS analysis contains a lot more data and differs in several ways from the above RGS work  \citep{2004ApJ...603..449C}. Yet, the results are consistent in the sense that neither detected the 
\ion{O}{VIII}$_{\rm A3}$ line, indicating no significant systematic uncertainties in either RGS results.

\subsubsection{Co-addition}
However, co-addition of a large number of spectra, 25 in our case, may be problematic considering weak unresolved lines.
In order to examine the robustness of our results we now repeated our \ion{O}{VIII}$_{\rm A3}$ analysis, but this time instead of co-adding the 25 RGS1 residual spectra, see \citet{2017A&A...605A..47N} for the information on the observation identification codes), we analysed them jointly. Here we started from the archival data and processed them with the standard procedures available in Oct 2017, differently from the procedure described in section~\ref{xray}.
In detail, we fitted the 19.05--20.2 \AA\ continua with a power-law model, allowing the normalisations to vary, independently from each other.
We adopted the galactic absorption model described in section~\ref{galabs}.
We added a slab component for \ion{O}{VIII} absorption redshifted by z$\equiv$0.055.
linking the column density equal in all spectra. The result is EW(\ion{O}{VIII}) $\le$ 0.9 m\AA , very similar to what we derived using the co-added RGS1 spectrum 
EW(\ion{O}{VIII}) $\le$ 1.3 m\AA .
Thus, we conclude that we have not missed a significant \ion{O}{VIII}~Ly$\alpha$ line at z~=~0.055 due to possible co-addition problems in the RGS1 data.
The close similarity of the two results also demonstrates that the non-standard processing of the RGS data we adopted for our work see \citet{2017A&A...605A..47N} for details) does not produce significant problems.

\subsection{LETG/ACIS issues}
\label{acis}
The fact that the LETG/ACIS is the only instrument combination that has yielded significant detections of \ion{O}{VIII}$_{\rm A3}$ indicates significant systematic uncertainties in the LETG/ACIS data.
Thus, we investigated next the hypothesis that the LETG/ACIS \ion{O}{VIII}$_{\rm A3}$ detections are actually due to a line-like artifact in the LETG/ACIS data as analysed by \citet{2007ApJ...670..992F} and \citet{2007ApJ...665..247W}.

\subsubsection{Aim point offset}
As discussed by  \citet{2007ApJ...670..992F} and \citet{2007ApJ...665..247W}, most of the observations they used have been obtained with an aim point offset of $\Delta$y = +1.5 arcmin. Such configuration places the tentative positive first order 20.0 \AA\ line close to the boundary between nodes 2 and 3 in the S3 chip. 
The standard dithering of the spacecraft (16 arcsec peak-to-peak i.e. $\sim$1 \AA ) will fill the gap with photons.
Depending on the completeness of the procedure, the calibration of the effective area at 20.0 \AA\ may be less accurate than at other wavelengths. 

\citet{2007ApJ...670..992F} presented following arguments against significant calibration artifacts at $\lambda \sim$ 20.0 \AA .
Firstly, the tentative negative first order $\lambda$ = 20.0 \AA\ line, when using the above offsets, is safely outside the node boundary. 
Yet, according to \citet{2007ApJ...670..992F} the line is clearly visible in both positive and negative first order data.
Unfortunately they did not report the wavelengths or EW measurements for these lines.
Secondly, the 20.0 \AA\ line obtained with observation 3669, which has been carried out with an exceptionally large offset ($\Delta$y = +3.3 arcmin), lands at a very different detector location compared to the other observations. 
However, this argument is not developed to the point of comparing the results obtained with this observation with the rest of the sample.

\subsubsection{Methods}

We examined the above arguments by re-analysing the LETG/ACIS data used by \citet{2007ApJ...670..992F} and \citet{2007ApJ...665..247W}.
The data used in the two works are almost the same, with the exceptions that the latter did not include the very high offset observation 3669 (42 ks), and they included an additional observation 3668 (14 ks). Also, \citet{2007ApJ...665..247W} did not shift the energy scales of the two observations as  \citet{2007ApJ...670..992F}. Yet, the two works yielded a significant detection of \ion{O}{VIII} at z $\approx$ 0.054--0.055 with consistent EW values.

We processed the data with the current public CIAO processing pipeline tool $chandra\_repro$,
and for co-adding the +1 and -1 order data we used the tool $combine\_grating\_spectra$.
We used  a bin size of 25 m\AA\ as in \citet{2007ApJ...670..992F}. 
The inclusion of the 2\% systematic uncertainties (see Section~\ref{xray}) had no significant effect because the statistical uncertainties are larger due to shorter exposure and lower effective area compared to that of RGS.

As in the case of RGS and HRC (see section~\ref{gallines}) we examined here the wavelength scale calibration accuracy of each LETG/ACIS co-added spectrum using Galactic lines.
We then fitted the 19-21 \AA\ band continuum with a spline model\footnote{In case of the \ion{O}{VIII}$_{\rm A3}$ line, the usage of the more complex continuum, including the Galactic absorbers as in section~\ref{galabs}, yielded a negligible effect on the  \ion{O}{VIII} equivalent width in the narrow band we used.}
with a grid step of 0.75 \AA\ to mimic the continuum modelling of 
\citet{2007ApJ...670..992F}, where the polynomial fits effectively removed features broader than 0.7 \AA .
We then added a redshifted \ion{O}{VIII} slab model to investigate the tentative WHIM line.

\subsubsection{Tests with \citet{2007ApJ...665..247W} sample}
\label{will} 
\begin{figure*}
\includegraphics[width=18cm,angle=0]{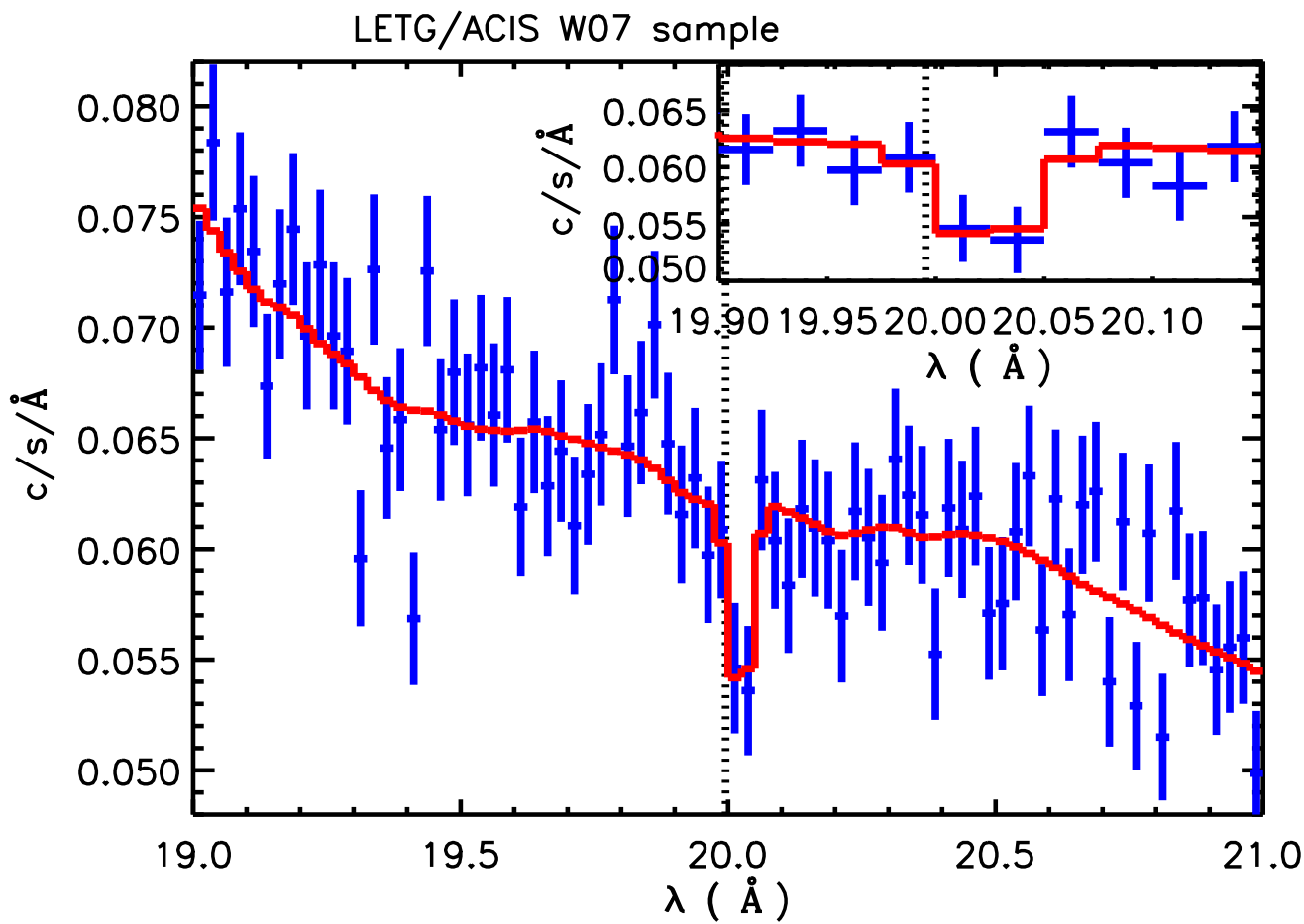}
\caption{The co-added LETG/ACIS data from \citet{2007ApJ...665..247W} observations (blue crosses), and the best-fit model (red line) consisting of a spline continuum
and a redshifted \ion{O}{VIII} doublet Voigt profile with the redshift as a free parameter. The dotted vertical line indicates the expected wavelength of \ion{O}{VIII} centroid, i.e. assuming the FUV redshift of z = 0.05405.
}
\label{wil_slab.fig}
\end{figure*}

The Galactic \ion{O}{I} and \ion{O}{VII} lines in our co-added \citet{2007ApJ...665..247W}  sample spectrum were unambiguously detected and indicated that the wavelength scale is very accurate, i.e. a possible shift is smaller than the statistical uncertainties of $\pm$6 m\AA , similarly as in \citet{2007ApJ...665..247W}.
The WHIM \ion{O}{VIII} modelling of the 19--21 \AA\ band data yielded a significant (3.6 $\sigma$) detection of a line at $\lambda$ = 20.03$\pm$0.01, identical with \citet{2007ApJ...665..247W} (see Fig.~\ref{wil_slab.fig}). Our EW = $8.2\pm2.3$ m\AA\ is consistent with \citet{2007ApJ...665..247W}  (EW = 7.5$\pm$2.1 m\AA ).

We then examined, whether the above detection may be an artifact produced by the node boundary (see above).
We utilised the argument that the tentative 20 \AA\ line in the positive and negative first order spectrum lands at a very different region of the detector.
Thus, possible consistence of the wavelengths and equivalent widths of the lines in both orders would indicate a true astrophysical origin for the line.
We used the same sample as \citep{2007ApJ...665..247W}, but this time separately for the positive and negative first order co-added spectra.

\begin{figure*}
\includegraphics[width=18cm,angle=0]{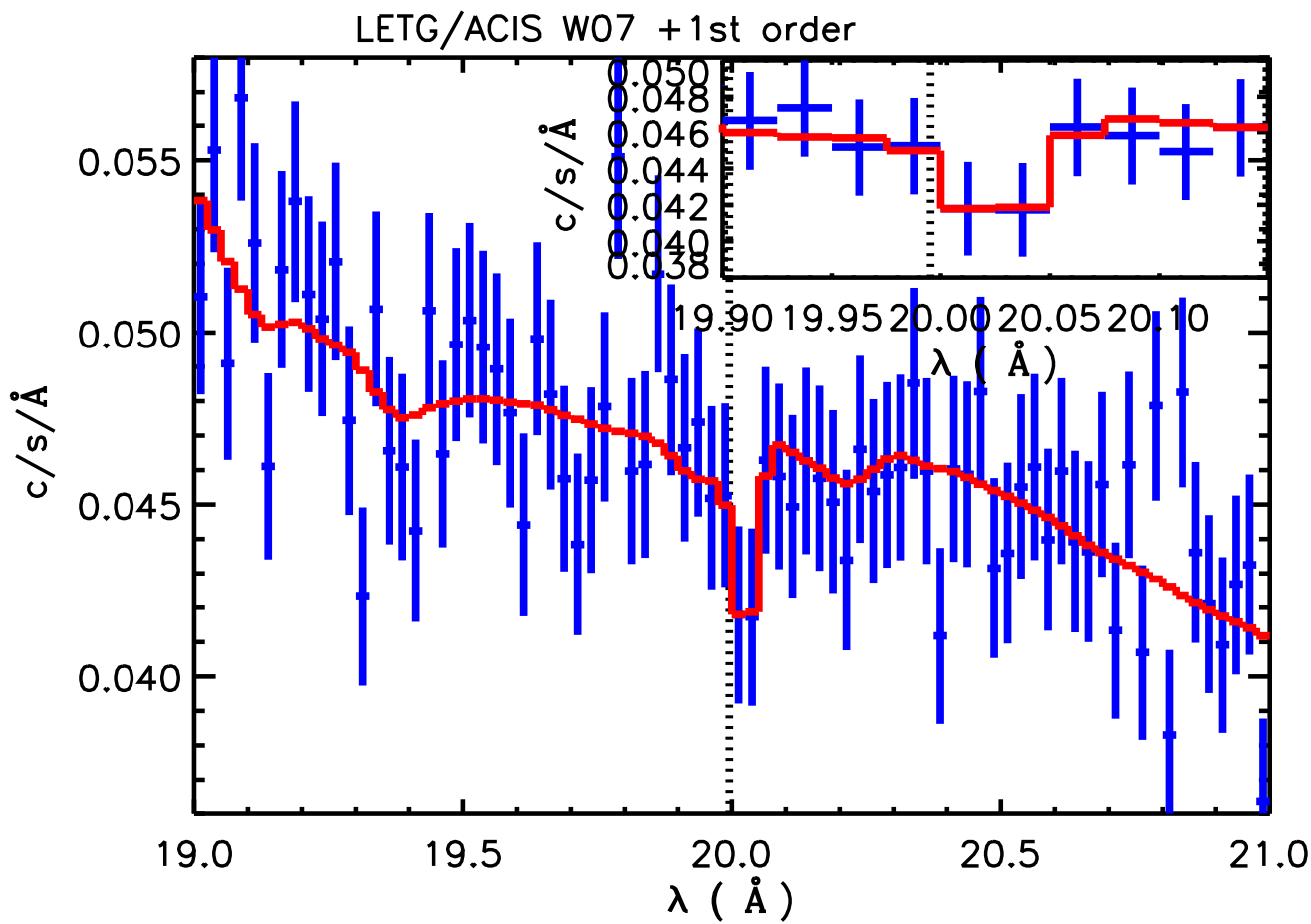}
\caption{ As Fig. \ref{wil_slab.fig} but using only positive 1st order data.
}
\label{wilp_slab.fig}
\end{figure*}

In the case of the positive order, the Galactic lines did not allow a significant wavelength scale shift.
Around 20 \AA\ the data indicated a line-like feature at $\lambda$ = 20.03$\pm$0.01, i.e. identical to that found when using the co-added positive and negative first order spectra above.
The equivalent width measurement of 6.5$\pm$2.7 m\AA\ yields a detection significance of 2.4 $\sigma$ (see Fig.~\ref{wilp_slab.fig}).

\begin{figure*}
\includegraphics[width=18cm,angle=0]{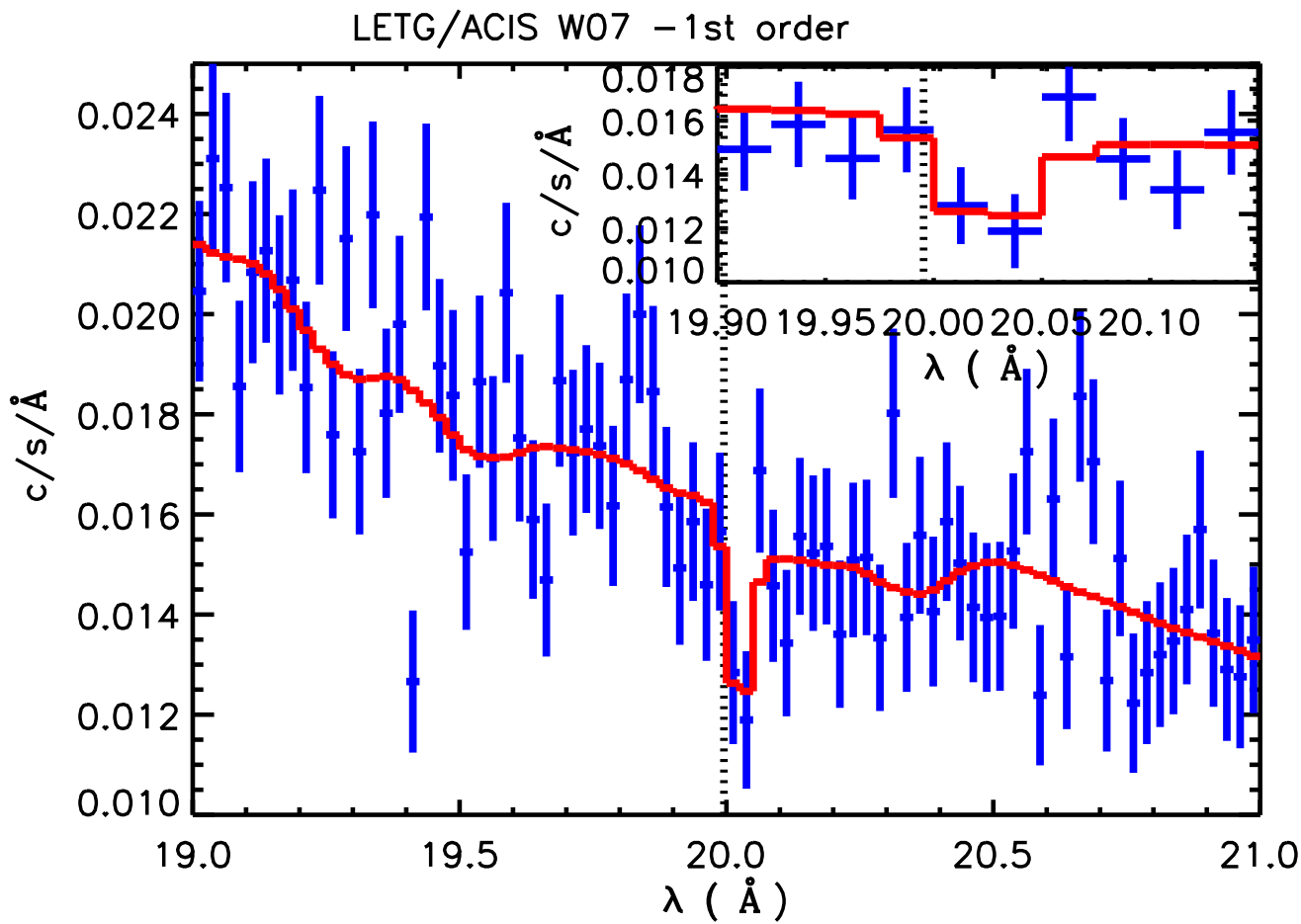}
\caption{ As Fig. \ref{wil_slab.fig} but using only negative 1st order data.
}
\label{wiln_slab.fig}
\end{figure*}

In case of the negative order, no Galactic line was unambiguously detected. 
At the vicinity of 20 \AA\ we detected an
absorption line at $\lambda$ = 20.02$\pm$0.01 \AA\,  consistently with the positive order
(see Fig.~\ref{wiln_slab.fig}).
We measured  EW = 13.4$\pm$5.0  m\AA\ for this line, (2.8 $\sigma$), consistently with the positive order. 

While the statistical significances of the positive and negative order lines are not very high individually,
their centroids are consistent within 1-3 $\sigma$ with that 
of \ion{O}{VIII}~Ly$\alpha$ (20.00 \AA ) at z=0.05425, the redshift of FUV absorber A3 (see Table~\ref{tab:absorbers}). 
This match will improve the probability of the line feature being a true astrophysical signal (see below).

\subsubsection{Tests with the large offset observation 3669}
\label{3669}
We then utilised the \citet{2007ApJ...670..992F} argument that the 20~\AA\ line in the large off-axis observation 3669 lands at a different detector region, compared to the above
lower offset sample case, and thus is not affected by the same node boundary. Thus it is very unlikely that random fluctuations would yield consistent lines 
using observation 3669 and the \citet{2007ApJ...665..247W} sample above.

The centroid of the Galactic \ion{O}{I} (\ion{O}{VII} was not useful) absorption line in observation 3669 was significantly shifted towards lower wavelengths by 40$\pm$10 m\AA .
Such a large wavelength shift may be due to complications of the calibration of large offset observations.

\begin{figure*}
\includegraphics[width=18cm,angle=0]{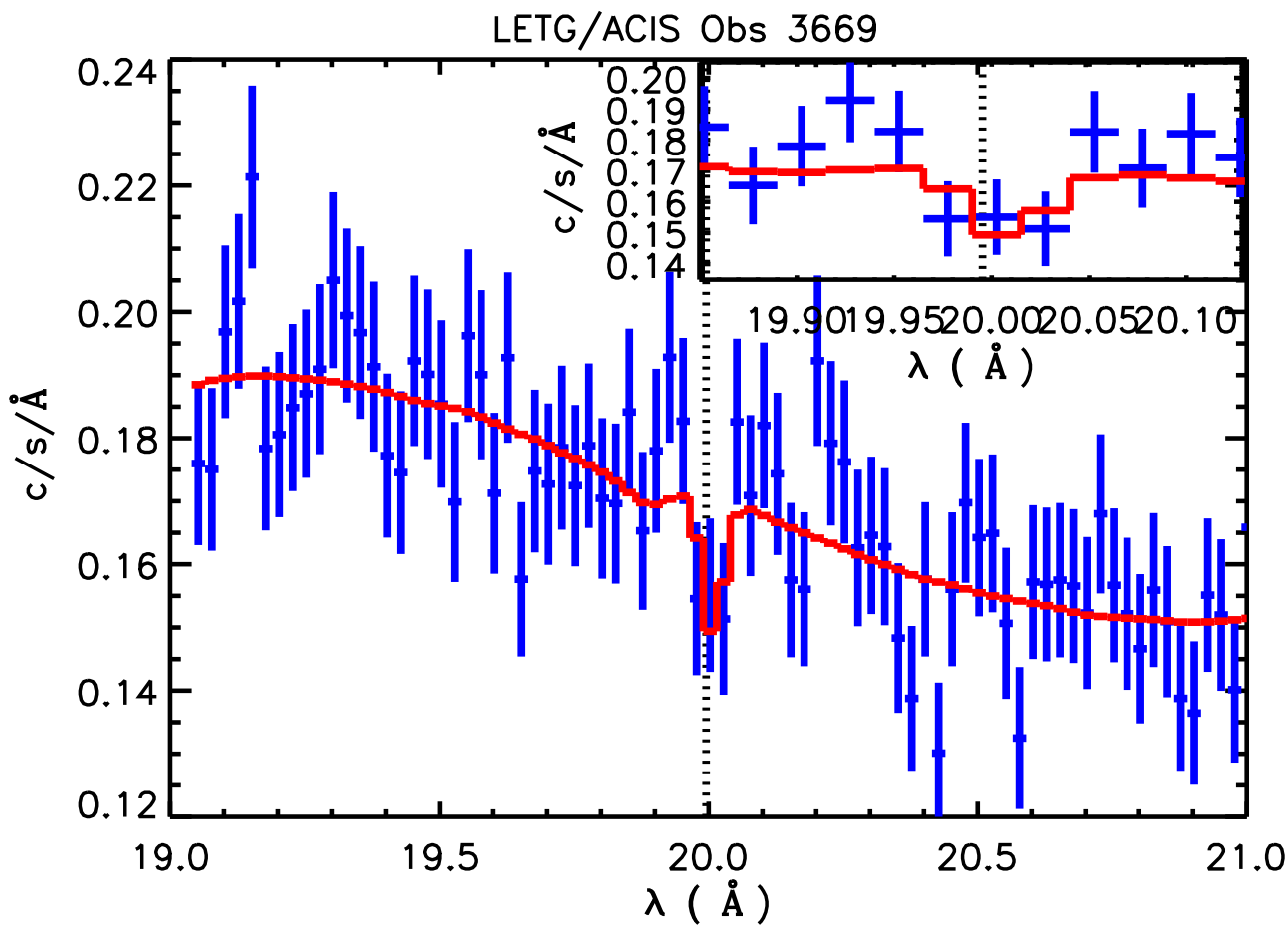}
\caption{As Fig. \ref{wil_slab.fig} but using only observation 3669. The wavelengths of the data and the best-fit model are shifted by 
+40 m\AA\ , as suggested by the Galactic \ion{O}{I} and \ion{O}{VII} lines. 
}
\label{3669_slab.fig}
\end{figure*}

At wavelengths around 20~\AA\  we found a 2.2 $\sigma$ indication for an absorption line at $\lambda = 19.97^{+0.01}_{-0.02}$ \AA\ (see Fig.~\ref{3669_slab.fig}).
Since 20~\AA\ photons land quite far from the FOV centre in such a high offset observation, the energy resolution is degraded.
This is seen via the broad, $\sim$75~m\AA\ wide feature in the data (see Fig.~\ref{3669_slab.fig}).
Due to the different amount of broadening at different offset angles, it is important not to co-add observations with different offsets.
The energy redistribution function shows a similarly broad feature, indicating that the off-set line spread function is well calibrated.  
Using the slab model we measured EW~=~$6.9\pm3.2$~m\AA\ for this line (no additional broadening required) which is consistent with our measurement using
the \citet{2007ApJ...665..247W} sample, see section~\ref{will}.
Shifting the line centroid by +40$\pm$10~m\AA , as indicated by the Galactic OI,
the corrected wavelength $\lambda = 20.01^{+0.01}_{-0.02}$~\AA\ of the \ion{O}{VIII} line is consistent with that derived above using the \citet{2007ApJ...665..247W} sample.

While the statistical significance of the line detection is quite low, its wavelength consistence with that of \ion{O}{VIII}~Ly$\alpha$ at the redshift of the FUV absorber A3 renders this test indicative that the line is a true astrophysical signal.

\begin{table}
 \centering
  \caption{Tests to LETG/ACIS data at $\lambda \sim$ 20\AA 
  \label{letg_tests.tab}}
    \begin{tabular}{lcc}
  \hline\hline
                &           &       \\ 
  sample   	& $\lambda$ & EW	\\
                & \AA       & m\AA  \\
  \hline 
W07$\pm$\tablefootmark{a}	        & 20.03$\pm$0.01 	& 8.2$\pm$2.3		\\
W07+\tablefootmark{b}	        & 20.03$\pm$0.01	& 6.5$\pm$2.7		\\
W07-\tablefootmark{c}	        & 20.02$\pm$0.01	& 13.4$\pm$5.0		\\
3669\tablefootmark{d}            & 20.01$^{+0.01}_{-0.02}$\tablefootmark{g}	& 6.9$\pm$3.2		\\
final\tablefootmark{e}  	 & 20.02$\pm$0.01	& 6.0$\pm$1.9		\\
final \& 3669 \tablefootmark{f}  & 20.02$\pm$0.01               & 6.2$\pm$1.7		\\
                &                \\
\hline                  
\end{tabular}
\tablefoot{\\
\tablefoottext{a}{The observations used in \citet{2007ApJ...665..247W}, positive and negative first orders co-added.} 
\tablefoottext{b}{The observations used in \citet{2007ApJ...665..247W}, but only positive first order.} 
\tablefoottext{c}{The observations used in \citet{2007ApJ...665..247W}, but only negative first order.} 
\tablefoottext{d}{The large pointing offset observation 3669.}
\tablefoottext{e}{Using all the public data in Oct 2017 obtained with $\Delta$y = +1.5 arcmin and SIM-Z = -8 mm.}
\tablefoottext{f}{Combining the results of the final sample and of the observation 3669.}
\tablefoottext{g}{After correcting the wavelength scale as indicated by the shift of the wavelength of the Galactic \ion{O}{I} line.}
}
\end{table}

\subsubsection{The final sample}
\label{final}
\begin{figure*}
\includegraphics[width=18cm,angle=0]{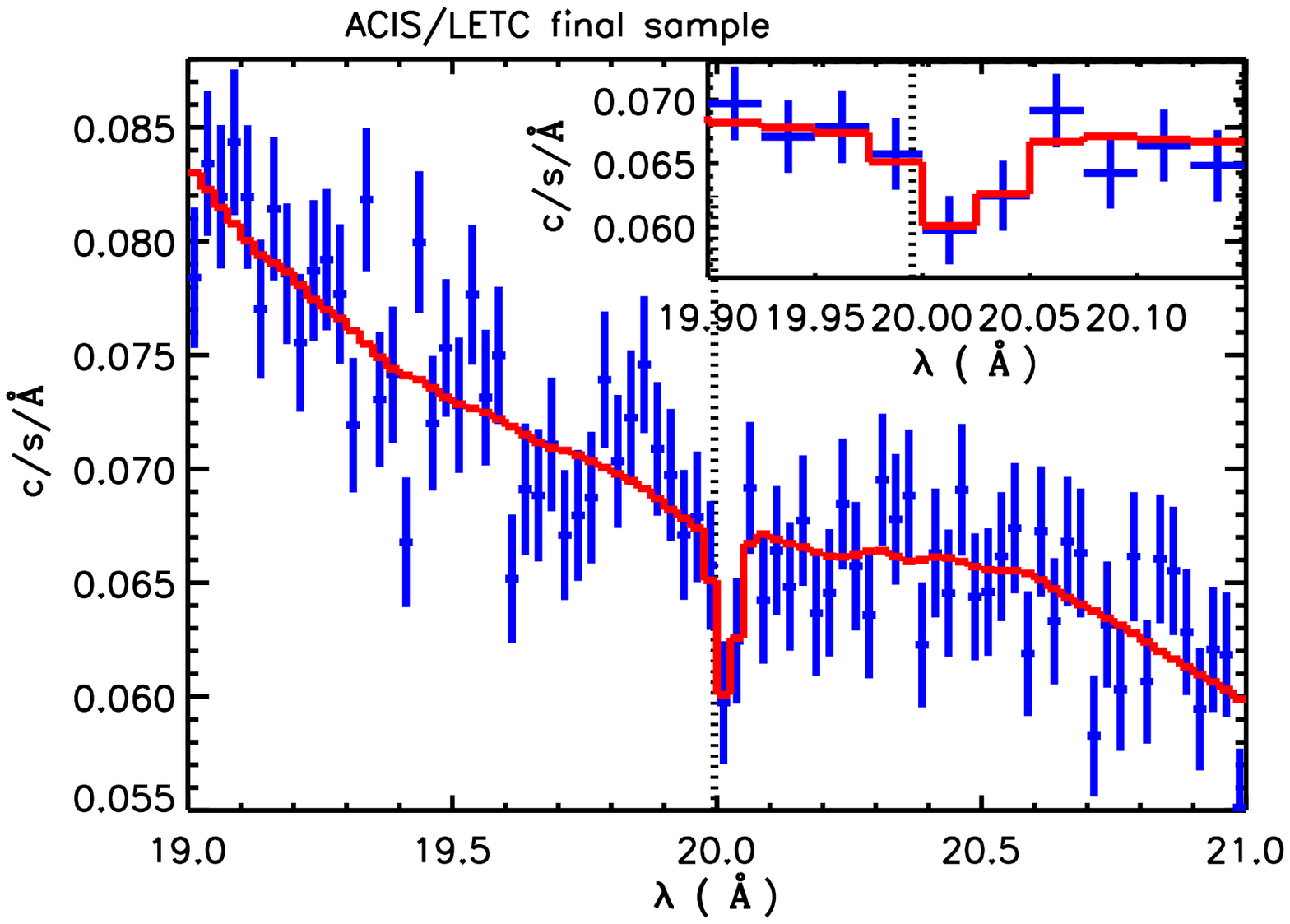}
\caption{As Fig. \ref{wil_slab.fig} but using the ``final'' sample.  
}
\label{final_slab.fig}
\end{figure*}

We then utilised the additional useful observations of \PKS\ obtained after the works published by 
\citet{2007ApJ...665..247W} and \citet{2007ApJ...670..992F}, available in Oct 2017 (Table~\ref{letg.tab}).
Due to the indicated problems with the large offset pointing wavelength scale calibration (see section~\ref{3669})
we selected only the observations obtained with the same offset setting as in the above works, i.e. $\Delta$y = +1.5 arcmin and SIM-Z = -8 mm.
Our co-added spectrum has an exposure time of 330 ks.
The study of the wavelengths of the Galactic \ion{O}{I} and \ion{O}{VII} lines allowed no wavelength scale shift.
We significantly detected a line at z=0.0554$\pm$0.0004 or $\lambda$ = 20.02$\pm$0.01 \AA\, (see Fig.~\ref{final_slab.fig}) with EW = 6.0$\pm$1.9 m\AA\
(3.2 $\sigma$), consistently with our analysis of the observation 3669 (see section~\ref{3669}).

We did not co-add the data from the observation 3669 to that of the final sample due to the indicated significant wavelength shift.
Since \PKS\ was very bright during the observation 3669, it has significant statistical weight compared to the final sample. 
We thus combined the measurements of the equivalent width using the spectra of the final sample and the observation 3669 into an error weighted average, 
obtaining EW = 6.2$\pm$1.7 for the $\lambda \approx $ 20 \AA\ line which is thus detected at a quite high statistical significance of 3.7 $\sigma$.

\begin{table}
 \centering
  \caption{The final sample of LETG/ACIS observations
  \label{letg.tab}}
    \begin{tabular}{cc}
  \hline\hline
                &                       \\ 
  OBS. ID.	&	Exposure (ks)	\\
  \hline 
	2335	&	29	\\
	3168	&	29	\\
	3668	&	13	\\
	3707	&	27	\\
	4416	&	46	\\
	6090	&	28	\\
	6091	&	29	\\
	6874	&	28	\\
	6924	&	9	\\
	6927	&	27	\\
	7293	&	9	\\
	8388	&	29	\\
        10662   &        28     \\
                &                \\
\hline                  
\end{tabular}
\end{table}

\subsection{Summary}

Our measurements of the positive and negative orders of the \citet{2007ApJ...665..247W} sample and the large offset observation 3669
yielded consistent wavelengths and equivalent widths for the \ion{O}{VIII}~Ly$\alpha$ line, indicating that the systematic uncertainties are small.
Thus, our tests indicated that the feature is very likely a true astrophysical signal.
The assumption that the ACIS measurement is true can be formulated as a null hypothesis that by change the ACIS measurement agrees with the non-detection of RGS1 ($\le$~0.9~m\AA ).
Applying the $\chi^{2}$ statistics to the ACIS analysis yielded a very low 0.7\% probability for such situation.
The random probability would become even smaller when considering the matches obtained with the above tests of positive and negative orders and the offset observation which cannot be derived analytically.

On the other hand, it is also very unlikely that $\sim$1.2~Ms of RGS1 data and $\sim$300~ks of HRC data could have missed a true \ion{O}{VIII}~Ly$\alpha$ line (see section~\ref{RGSissues})
with EW~=~6~m\AA , as measured with ACIS. The application of the $\chi^{2}$ statistics to the RGS1 analysis yielded that such incidence has a 0.006\% probability.

Complicated and extensive simulations would be needed to improve the accuracy of the above probabilities. 
However, we think that this is not useful since both values would be extremely high, and thus we could not
conclude which case is significantly better. Thus, given the very convincing X-ray spectral evidence for and against the existence of the $\lambda \sim$ 20~\AA\ absorption line, we cannot conclude whether or not the feature is a true astrophysical line.

\section{Discussion}
\label{impli}

\subsection{About the spatial co-incidence of \ion{O}{VII} absorbers at the location of \ion{O}{VI} absorbers}
\label{o6o7}
In this work we assumed the co-location of the warm and hot WHIM in order to enable the usage of the robustly detected warm WHIM in the FUV band as a tracer of the cosmologically interesting hot WHIM. We performed an X-ray follow-up of the \ion{O}{VII} line in the \PKS\ sight line at the two redshifts where the previous FUV observations \citep{2012ApJ...759..112T} have yielded \ion{O}{VI} detections at the level of  $\log{N(\ion{O}{VI}({\rm cm}^{-2}))}~\sim~14$.
We obtained no significant detection and upper (RGS1) limits of $\log{N(\ion{O}{VII}({\rm cm}^{-2}))}~\le$~14.5 and 15.2\footnote{The value is higher due to complications with the overlapping Galactic \ion{O}{IV} line} for the two absorbers, i.e. A3 and A4 (see Table~\ref{results.tab} and Fig.~\ref{WHIM_OVII_RGS1.fig}).

Considering the simulations of \citet{2012ApJ...753...17C} the basic problem in our work is the shortness of the studied redshift path, dz=0.1.
The above simulations indicated about one \ion{O}{VI} absorber with $\log{N(\ion{O}{VI}({\rm cm}^{-2}))}~\sim~14$
and one \ion{O}{VII} absorber at the level of $\log{N(\ion{O}{VII}({\rm cm}^{-2}))}~\sim~15$ within our path length on average.
The observational values for the above absorbers are two (\ion{O}{VI}) and zero (\ion{O}{VII}), similar to the prediction.
Additionally, in the above simulation \citep{2012ApJ...753...17C}, the probability of finding an absorber with $\log{N(\ion{O}{VII}({\rm cm}^{-2}))}~\sim~15$ 
at the location of an absorber with $\log{N(\ion{O}{VI}({\rm cm}^{-2}))}~\sim~14$ is less than 10\%.
Thus, we should increase the studied path length at least by an order of magnitude to get a meaningful sample for comparisons with the simulations.
As we demonstrated in this work, the study of the \ion{O}{VII} line at even a small path length at the level $\log{N(\ion{O}{VII}({\rm cm}^{-2}))}~\sim~15$
is very challenging.
The requirements for the signal-to-noise with the current instruments translate into very long exposures of several 100~ks.
We plan to probe systematically the current archival high resolution X-ray spectra at the locations of the FUV-detected warm WHIM in a future work.

\subsection{LETC/ACIS $\lambda~\sim$~20~\AA\ feature}
The LETG/ACIS feature at $\lambda~\sim$~20~\AA\ in our final sample (see Fig.~\ref{final_slab.fig}), if interpreted as being due to \ion{O}{VIII} at absorber A3 (z$\sim0.054$), corresponds to $\log{N(\ion{O}{VIII}({\rm cm}^{-2}))}~=~ 15.7^{+0.1}_{-0.2}$. At consistent redshifts there is very securely detected FUV absorption (see Table~\ref{tab:absorbers}) yielding  $\log{N(\ion{O}{VI}({\rm cm}^{-2}))}~=~13.6\pm0.1$ \citep{2012ApJ...759..112T}. 
The simulations of \citet{2012ApJ...753...17C} indicate that in such a small redshift path as studied in our work (dz$\sim$0.1) there should be on average
only $\sim$0.1 \ion{O}{VII} absorbers with column density exceeding $10^{15.7}$~cm$^{-2}$.
Since the IllustrisTNG simulations \citep{2017arXiv171200016N} indicate that clustering of \ion{O}{VIII}  ions in the Mpc scales is an order of magnitude smaller than that of \ion{O}{VII}  ions, the probability for finding \ion{O}{VIII} absorbers is much smaller than that ($\lesssim$~10\%) for \ion{O}{VII} discussed above in Section~\ref{o6o7}.
Thus, the probability of LETG/ACIS $\lambda~\sim~20$~\AA\ feature being due to astrophysical \ion{O}{VIII} absorber co-located with the FUV-detected \ion{O}{VI} absorber is at the very low level level of $\lesssim$~0.1\%.

Assuming that the LETG/ACIS $\lambda~\sim~20$~\AA\ feature is due to \ion{O}{VIII}, 
the absence of significant \ion{O}{VII} absorption at A3 indicates two distinct gas phases: the FUV-detected warm one with $\log{T({\rm K})}~\lesssim~6$ and the X-ray-detected hot one with $\log{T({\rm K})}~\gtrsim~6$.
Such configuration is possible in the WHIM embedded in a large scale filament. 
Namely, the cosmological simulations e.g. EAGLE \citep{2015MNRAS.446..521S} and IllustrisTNG \citep{2017arXiv171200016N} suggest that the hottest WHIM is concentrated along filament axes while the warm WHIM occupies much larger surrounding volumes.
Thus, most of the random sight lines are expected to contain only the warm FUV WHIM, consistent with the simulations of \citet{2012ApJ...753...17C}.
We suggest that in few lucky incidences (as may be the case for A3, as well as for the Sculptor Wall \citep{2009ApJ...695.1351B, 2010ApJ...714.1715F} and 3C273 (J. Ahoranta et al., 2018, in preparation)) when the sight line passes very close to the filament axis, both warm and hot WHIM may be co-located and thus the FUV 
and X-ray spectra would exhibit WHIM absorption lines at similar redshifts. 

Assuming that a) LETG/ACIS $\lambda~\sim~20$~\AA\ feature is due to \ion{O}{VIII} embedded in the WHIM in an intervening large scale filament, 
b) the metal abundances in the A3 absorber are 0.1 Solar, c) the temperature of the WHIM is quite high $10^{6.5}$~K\footnote
{If the temperature is higher, i.e. less optimal for \ion{O}{VIII}  production, the hydrogen density and thus the path length would become larger.} to keep the \ion{O}{VII} column density below the detection limit of $10^{15}$~cm$^{-2}$
and d) that the system is in collisional ionisation equilibrium, the LETG/ACIS measurement $\log{N(\ion{O}{VIII}({\rm cm}^{-2}))}~=~ 15.7^{+0.1}_{-0.2}$ corresponds to equivalent hydrogen column density level of $10^{20}$~cm$^{-3}$.  
Such a hot model contains also lines from \ion{Ne}{IX} (rest frame $\lambda~\sim$~13.447~\AA ) 
and \ion{Fe}{XVII} (rest frame $\lambda~\sim$~15.014~\AA ). The  \ion{Ne}{IX} line is consistent with the ACIS data. 
The \ion{Fe}{XVII} line is slightly overpredicted with such a model, but within the statistical uncertainties, the iron abundance is consistent with 0.1 Solar used for oxygen.
Thus, the CIE modelling did not rule out the  \ion{O}{VIII} line.
 
Adopting the typical baryon overdensity range of 10-100 for the WHIM implies that the path through the WHIM should be in excess of 10~Mpc.
This in turn requires a co-incidence of a major filament being oriented very closely along the sight line.
We plan to address the puzzle of the A3 \ion{O}{VIII} line in a future work by utilising the galaxy distribution around the possible X-ray absorber in order to detect or rule out a major galactic filament crossing the \PKS\ sight line at the matching redshift.

\subsection{Transient LETG/ACIS $\lambda~\sim$~20~\AA\ absorption?}
In the case of the blazar H2356-309,  \citet{2011ApJ...731...46F} detected a transient \ion{O}{VIII} absorption feature (duration $\approx$~100~ks) at the surface of the blazar using LETG/HRC.
The feature was found consistent with thermal instability of the absorber and with an outflow of absorbing material.
We examine here whether a similar scenario could explain the LETG/ACIS detection of the line feature at $\lambda~\sim$~20~\AA\ in the case of \PKS .

LETC/ACIS observation 3669 (not included in the \citet{2007ApJ...665..247W} sample, see Section~\ref{3669}) and the \citet{2007ApJ...665..247W} sample (Section~\ref{will}) both show the line consistently.
This can be interpreted as two distinct transients with similar strength. 
It is not very likely that one or two transient absorption events took place when LETG/ACIS was observing \PKS\ , and none
took place when RGS or LETG/HRC observed \PKS . Yet the epoch sampling of \PKS\ with RGS and LETG is not frequent enough to rule out this possibility completely.

If the X-ray feature at $\lambda~\sim$~20~\AA , measured with LETG/ACIS only, is due to \ion{O}{VIII} at the blazar surface (z~$\approx$~0.12), the outflow velocity is dz $\times$  c $~\approx$ (1.12 - 1.05) $\times$ c $~\approx$~20000~km~s$^{-1}$ towards us. 
Such ultra-fast outflows (UFOs)  have been observed in the X-ray spectra of many Seyfert
galaxies (e.g. \citet{2014MNRAS.443L.104T}). However, we are not aware of any detection of similarly fast outflow in a blazar, such as \PKS\ .

Let's assume that blazars have not been sufficiently surveyed to robustly rule out the UFO scenario for \PKS .
However, the LETG/ACIS X-ray feature is too narrow and too weak compared to absorption lines measured in UFOs related to Seyferts (e.g. \citet{2014MNRAS.443L.104T}). Furthermore, in the UFO scenario the outflow at the blazar surface happens to have such a velocity that its combined effect with the Hubble velocity redshifts the \ion{O}{VIII} line to $\lambda~\approx$~20~\AA , which co-incides with the wavelength of \ion{O}{VIII} if originating from an absorber with only Hubble velocity at the location of the FUV absorber A3 (\ion{O}{VI} and two BLAs, see Table\ref{tab:absorbers}).

In summary, given the above problems, it is very unlikely that the transient absorption explains the LETG/ACIS $\lambda~\sim$~20~\AA\ feature.

\section{Conclusions}
We analysed all available useful high-resolution X-ray spectral data in the direction of the blazar \PKS\ at 
the redshifts where FUV observations \citep{2012ApJ...759..112T} have obtained indication of the warm WHIM. 
The FUV measurements consist of two absorbers with  $\log{N(\ion{O}{VI}({\rm cm}^{-2}))}~=~13.6\pm0.1$, 
one with  $\log{N(\ion{Si}{IV}({\rm cm}^{-2}))}~\sim~12.1\pm0.1$, 
and several BLA:s (\ion{H}{I} broadened  by b~$\ge$~40~km~s$^{-1}$) at the level of $\log{N(\ion{H}{I}({\rm cm}^{-2}))}~\sim~12-14$
(see Table~\ref{tab:absorbers}).
The studied redshift path is dz$\sim$0.1. The analysis yielded the following conclusions:

\begin{itemize}

\item
We did not obtain any significant detections of \ion{O}{VII}~He$\alpha$ absorption lines, the most likely hot WHIM tracer,
at the five (considering the \ion{O}{VI} lines and BLAs) or two (considering only the \ion{O}{VI} lines) FUV-based X-ray follow-up redshifts.
The non-detections correspond to upper RGS1 limits of $\log{N(\ion{O}{VII}({\rm cm}^{-2}))}~\le$~14.5-15.2

\item
At five of the six studied redshifts we did not detect any significant \ion{O}{VIII}~Ly$\alpha$ absorption line.
The upper limit is $\log{N(\ion{O}{VIII}({\rm cm}^{-2}))}~\lesssim~15$.

\item
The LETG/ACIS combination yielded an significant (3.7$\sigma$) detection of an absorption line - like feature at  $\lambda$~$\sim$~20~\AA .
If interpreting this as a true absorption line due to \ion{O}{VIII}, its redshift matches one of the six FUV-based X-ray follow-up redshifts (z$\sim$~0.054).

\item
The data from RGS1 and LETG/HRC did not detect the LETG/ACIS $\lambda$~$\sim$~20~\AA\ feature.
Given the high statistical quality of the RGS1 data, the possibility of RGS1 accidentally missing the true line at $\lambda$~20~\AA\
is very low,  0.006\%. 

\item
  Considering the simulations \citep{2012ApJ...753...17C} and \citep{2017arXiv171200016N}, the probability of LETG/ACIS $\lambda~\sim~20$~\AA\ feature being due to astrophysical \ion{O}{VIII} absorber co-located with the FUV-detected \ion{O}{VI} absorber is at the very low level level of $\lesssim$~0.1\%.

\item
We cannot rule out completely the very unlikely possibility that the  LETG/ACIS 20~\AA\ feature is due to a transient event located close to the blazar.

\end{itemize}

\begin{acknowledgements}
  We acknowledge the support by the Estonian Research Council grants PUT246, IUT26-2,
  IUT40-2, and by the European Regional Development Fund (TK133 and MOBTP86). Thanks to the Chandra X-ray observatory HelpDesk. JN acknowledges the funds from 
  a European Horizon 2020 program AHEAD (Integrated Activities in the High Energy Astrophysics Domain), and from FINCA (the Finnish Centre for Astronomy with 
ESO). Thanks to Jelle de Plaa for his help with the Spex analysis. 
  \end{acknowledgements}

\bibliographystyle{aa} 
\bibliography{pkswhimbib} 

\begin{thebibliography}{40}
\expandafter\ifx\csname natexlab\endcsname\relax\def\natexlab#1{#1}\fi

\bibitem[{{Branchini} {et~al.}(2009){Branchini}, {Ursino}, {Corsi}, {Martizzi},
  {Amati}, {den Herder}, {Galeazzi}, {Gendre}, {Kaastra}, {Moscardini},
  {Nicastro}, {Ohashi}, {Paerels}, {Piro}, {Roncarelli}, {Takei}, \&
  {Viel}}]{2009ApJ...697..328B}
{Branchini}, E., {Ursino}, E., {Corsi}, A., {et~al.} 2009, \apj, 697, 328

\bibitem[{{Buote} {et~al.}(2009){Buote}, {Zappacosta}, {Fang}, {Humphrey},
  {Gastaldello}, \& {Tagliaferri}}]{2009ApJ...695.1351B}
{Buote}, D.~A., {Zappacosta}, L., {Fang}, T., {et~al.} 2009, \apj, 695, 1351

\bibitem[{{Cagnoni} {et~al.}(2004){Cagnoni}, {Nicastro}, {Maraschi}, {Treves},
  \& {Tavecchio}}]{2004ApJ...603..449C}
{Cagnoni}, I., {Nicastro}, F., {Maraschi}, L., {Treves}, A., \& {Tavecchio}, F.
  2004, \apj, 603, 449

\bibitem[{{Cen}(2012)}]{2012ApJ...753...17C}
{Cen}, R. 2012, \apj, 753, 17

\bibitem[{{Cen} \& {Ostriker}(1999)}]{1999ApJ...514....1C}
{Cen}, R. \& {Ostriker}, J.~P. 1999, \apj, 514, 1

\bibitem[{{Cui} {et~al.}(2012){Cui}, {Borgani}, {Dolag}, {Murante}, \&
  {Tornatore}}]{2012MNRAS.423.2279C}
{Cui}, W., {Borgani}, S., {Dolag}, K., {Murante}, G., \& {Tornatore}, L. 2012,
  \mnras, 423, 2279

\bibitem[{{Cui} {et~al.}(2018){Cui}, {Knebe}, {Yepes}, {Yang}, {Borgani},
  {Kang}, {Power}, \& {Staveley-Smith}}]{2018MNRAS.473...68C}
{Cui}, W., {Knebe}, A., {Yepes}, G., {et~al.} 2018, \mnras, 473, 68

\bibitem[{{Danforth} \& {Shull}(2008)}]{2008ApJ...679..194D}
{Danforth}, C.~W. \& {Shull}, J.~M. 2008, \apj, 679, 194

\bibitem[{{Danforth} {et~al.}(2016){Danforth}, {Stocke}, {France}, {Begelman},
  \& {Perlman}}]{2016ApJ...832...76D}
{Danforth}, C.~W., {Stocke}, J.~T., {France}, K., {Begelman}, M.~C., \&
  {Perlman}, E. 2016, \apj, 832, 76

\bibitem[{{Dav{\'e}} {et~al.}(2001){Dav{\'e}}, {Cen}, {Ostriker}, {Bryan},
  {Hernquist}, {Katz}, {Weinberg}, {Norman}, \& {O'Shea}}]{2001ApJ...552..473D}
{Dav{\'e}}, R., {Cen}, R., {Ostriker}, J.~P., {et~al.} 2001, \apj, 552, 473

\bibitem[{{de Vries} {et~al.}(2015){de Vries}, {den Herder}, {Gabriel},
  {Gonzalez-Riestra}, {Ibarra}, {Kaastra}, {Pollock}, {Raassen}, \&
  {Paerels}}]{2015A&A...573A.128D}
{de Vries}, C.~P., {den Herder}, J.~W., {Gabriel}, C., {et~al.} 2015, \aap,
  573, A128

\bibitem[{{de Vries} {et~al.}(2003){de Vries}, {den Herder}, {Kaastra},
  {Paerels}, {den Boggende}, \& {Rasmussen}}]{2003A&A...404..959D}
{de Vries}, C.~P., {den Herder}, J.~W., {Kaastra}, J.~S., {et~al.} 2003, \aap,
  404, 959

\bibitem[{{Dolag} {et~al.}(2006){Dolag}, {Meneghetti}, {Moscardini}, {Rasia},
  \& {Bonaldi}}]{2006MNRAS.370..656D}
{Dolag}, K., {Meneghetti}, M., {Moscardini}, L., {Rasia}, E., \& {Bonaldi}, A.
  2006, \mnras, 370, 656

\bibitem[{{Fang} {et~al.}(2011){Fang}, {Buote}, {Humphrey}, \&
  {Canizares}}]{2011ApJ...731...46F}
{Fang}, T., {Buote}, D.~A., {Humphrey}, P.~J., \& {Canizares}, C.~R. 2011,
  \apj, 731, 46

\bibitem[{{Fang} {et~al.}(2010){Fang}, {Buote}, {Humphrey}, {Canizares},
  {Zappacosta}, {Maiolino}, {Tagliaferri}, \&
  {Gastaldello}}]{2010ApJ...714.1715F}
{Fang}, T., {Buote}, D.~A., {Humphrey}, P.~J., {et~al.} 2010, \apj, 714, 1715

\bibitem[{{Fang} {et~al.}(2007){Fang}, {Canizares}, \&
  {Yao}}]{2007ApJ...670..992F}
{Fang}, T., {Canizares}, C.~R., \& {Yao}, Y. 2007, \apj, 670, 992

\bibitem[{{Fang} {et~al.}(2002){Fang}, {Marshall}, {Lee}, {Davis}, \&
  {Canizares}}]{2002ApJ...572L.127F}
{Fang}, T., {Marshall}, H.~L., {Lee}, J.~C., {Davis}, D.~S., \& {Canizares},
  C.~R. 2002, \apjl, 572, L127

\bibitem[{{Kaastra} {et~al.}(1996){Kaastra}, {Mewe}, \&
  {Nieuwenhuijzen}}]{1996uxsa.conf..411K}
{Kaastra}, J.~S., {Mewe}, R., \& {Nieuwenhuijzen}, H. 1996, in UV and X-ray
  Spectroscopy of Astrophysical and Laboratory Plasmas, ed. K.~{Yamashita} \&
  T.~{Watanabe}, 411--414

\bibitem[{{Lehner} {et~al.}(2007){Lehner}, {Savage}, {Richter}, {Sembach},
  {Tripp}, \& {Wakker}}]{2007ApJ...658..680L}
{Lehner}, N., {Savage}, B.~D., {Richter}, P., {et~al.} 2007, \apj, 658, 680

\bibitem[{{Lodders} \& {Palme}(2009)}]{2009M&PSA..72.5154L}
{Lodders}, K. \& {Palme}, H. 2009, Meteoritics and Planetary Science
  Supplement, 72, 5154

\bibitem[{{Nelson} {et~al.}(2017){Nelson}, {Kauffmann}, {Pillepich}, {Genel},
  {Springel}, {Pakmor}, {Hernquist}, {Weinberger}, {Torrey}, {Vogelsberger}, \&
  {Marinacci}}]{2017arXiv171200016N}
{Nelson}, D., {Kauffmann}, G., {Pillepich}, A., {et~al.} 2017, ArXiv e-prints
  [\eprint[arXiv]{1712.00016}]

\bibitem[{{Nevalainen} {et~al.}(2015){Nevalainen}, {Tempel}, {Liivam{\"a}gi},
  {Branchini}, {Roncarelli}, {Giocoli}, {Hein{\"a}m{\"a}ki}, {Saar}, {Tamm},
  {Finoguenov}, {Nurmi}, \& {Bonamente}}]{2015A&A...583A.142N}
{Nevalainen}, J., {Tempel}, E., {Liivam{\"a}gi}, L.~J., {et~al.} 2015, \aap,
  583, A142

\bibitem[{{Nevalainen} {et~al.}(2017){Nevalainen}, {Wakker}, {Kaastra},
  {Bonamente}, {Snowden}, {Paerels}, \& {de Vries}}]{2017A&A...605A..47N}
{Nevalainen}, J., {Wakker}, B., {Kaastra}, J., {et~al.} 2017, \aap, 605, A47

\bibitem[{{Nicastro} {et~al.}(2018){Nicastro}, {Kaastra}, {Krongold},
  {Borgani}, {Branchini}, {Cen}, {Dadina}, {Danforth}, {Elvis}, {Fiore},
  {Gupta}, {Mathur}, {Mayya}, {Paerels}, {Piro}, {Rosa-Gonzalez}, {Schaye},
  {Shull}, {Torres-Zafra}, {Wijers}, \& {Zappacosta}}]{2018Natur.558..406N}
{Nicastro}, F., {Kaastra}, J., {Krongold}, Y., {et~al.} 2018, \nat, 558, 406

\bibitem[{{Nicastro} {et~al.}(2016){Nicastro}, {Senatore}, {Gupta}, {Mathur},
  {Krongold}, {Elvis}, \& {Piro}}]{2016MNRAS.458L.123N}
{Nicastro}, F., {Senatore}, F., {Gupta}, A., {et~al.} 2016, \mnras, 458, L123

\bibitem[{{Richter} {et~al.}(2017){Richter}, {Nuza}, {Fox}, {Wakker}, {Lehner},
  {Ben Bekhti}, {Fechner}, {Wendt}, {Howk}, {Muzahid}, {Ganguly}, \&
  {Charlton}}]{2017A&A...607A..48R}
{Richter}, P., {Nuza}, S.~E., {Fox}, A.~J., {et~al.} 2017, \aap, 607, A48

\bibitem[{{Savage} {et~al.}(2014){Savage}, {Kim}, {Wakker}, {Keeney}, {Shull},
  {Stocke}, \& {Green}}]{2014ApJS..212....8S}
{Savage}, B.~D., {Kim}, T.-S., {Wakker}, B.~P., {et~al.} 2014, \apjs, 212, 8

\bibitem[{{Schaye} {et~al.}(2015){Schaye}, {Crain}, {Bower}, {Furlong},
  {Schaller}, {Theuns}, {Dalla Vecchia}, {Frenk}, {McCarthy}, {Helly},
  {Jenkins}, {Rosas-Guevara}, {White}, {Baes}, {Booth}, {Camps}, {Navarro},
  {Qu}, {Rahmati}, {Sawala}, {Thomas}, \& {Trayford}}]{2015MNRAS.446..521S}
{Schaye}, J., {Crain}, R.~A., {Bower}, R.~G., {et~al.} 2015, \mnras, 446, 521

\bibitem[{{Sembach} {et~al.}(2003){Sembach}, {Wakker}, {Savage}, {Richter},
  {Meade}, {Shull}, {Jenkins}, {Sonneborn}, \& {Moos}}]{2003ApJS..146..165S}
{Sembach}, K.~R., {Wakker}, B.~P., {Savage}, B.~D., {et~al.} 2003, \apjs, 146,
  165

\bibitem[{{Shull} {et~al.}(1998){Shull}, {Penton}, {Stocke}, {Giroux}, {van
  Gorkom}, {Lee}, \& {Carilli}}]{1998AJ....116.2094S}
{Shull}, J.~M., {Penton}, S.~V., {Stocke}, J.~T., {et~al.} 1998, \aj, 116, 2094

\bibitem[{{Shull} {et~al.}(2012){Shull}, {Smith}, \&
  {Danforth}}]{2012ApJ...759...23S}
{Shull}, J.~M., {Smith}, B.~D., \& {Danforth}, C.~W. 2012, \apj, 759, 23

\bibitem[{{Shull} {et~al.}(2003){Shull}, {Tumlinson}, \&
  {Giroux}}]{2003ApJ...594L.107S}
{Shull}, J.~M., {Tumlinson}, J., \& {Giroux}, M.~L. 2003, \apjl, 594, L107

\bibitem[{{Stocke} {et~al.}(2013){Stocke}, {Keeney}, {Danforth}, {Shull},
  {Froning}, {Green}, {Penton}, \& {Savage}}]{2013ApJ...763..148S}
{Stocke}, J.~T., {Keeney}, B.~A., {Danforth}, C.~W., {et~al.} 2013, \apj, 763,
  148

\bibitem[{{Stocke} {et~al.}(2014){Stocke}, {Keeney}, {Danforth}, {Syphers},
  {Yamamoto}, {Shull}, {Green}, {Froning}, {Savage}, {Wakker}, {Kim},
  {Ryan-Weber}, \& {Kacprzak}}]{2014ApJ...791..128S}
{Stocke}, J.~T., {Keeney}, B.~A., {Danforth}, C.~W., {et~al.} 2014, \apj, 791,
  128

\bibitem[{{Tilton} {et~al.}(2012){Tilton}, {Danforth}, {Shull}, \&
  {Ross}}]{2012ApJ...759..112T}
{Tilton}, E.~M., {Danforth}, C.~W., {Shull}, J.~M., \& {Ross}, T.~L. 2012,
  \apj, 759, 112

\bibitem[{{Tombesi} \& {Cappi}(2014)}]{2014MNRAS.443L.104T}
{Tombesi}, F. \& {Cappi}, M. 2014, \mnras, 443, L104

\bibitem[{{Wakker} {et~al.}(2003){Wakker}, {Savage}, {Sembach}, {Richter},
  {Meade}, {Jenkins}, {Shull}, {Ake}, {Blair}, {Dixon}, {Friedman}, {Green},
  {Green}, {Kruk}, {Moos}, {Murphy}, {Oegerle}, {Sahnow}, {Sonneborn},
  {Wilkinson}, \& {York}}]{2003ApJS..146....1W}
{Wakker}, B.~P., {Savage}, B.~D., {Sembach}, K.~R., {et~al.} 2003, \apjs, 146,
  1

\bibitem[{{Williams} {et~al.}(2007){Williams}, {Mathur}, {Nicastro}, \&
  {Elvis}}]{2007ApJ...665..247W}
{Williams}, R.~J., {Mathur}, S., {Nicastro}, F., \& {Elvis}, M. 2007, \apj,
  665, 247

\bibitem[{{Williams} {et~al.}(2013){Williams}, {Mulchaey}, \&
  {Kollmeier}}]{2013ApJ...762L..10W}
{Williams}, R.~J., {Mulchaey}, J.~S., \& {Kollmeier}, J.~A. 2013, \apjl, 762,
  L10

\bibitem[{{Yao} {et~al.}(2009){Yao}, {Tripp}, {Wang}, {Danforth}, {Canizares},
  {Shull}, {Marshall}, \& {Song}}]{2009ApJ...697.1784Y}
{Yao}, Y., {Tripp}, T.~M., {Wang}, Q.~D., {et~al.} 2009, \apj, 697, 1784

\end{thebibliography}

\end{document}